\documentclass[aps,floatfix,amsmath,nofootinbib,amssymb,towcolumn,superscriptaddress]{revtex4-2}

\usepackage{overpic}
\usepackage{amssymb}
\usepackage{indentfirst}
\usepackage{feynmf}   %{feynmp}
\usepackage{slashed}  %for Feynman symbols
\usepackage{cases}
\usepackage{color}
\usepackage{multirow}
\usepackage{epstopdf}
\usepackage{graphicx,color,bm}
\usepackage{epstopdf}

\usepackage[colorlinks,
            citecolor=green,
            anchorcolor=red,
            menucolor=red,
            linkcolor=red,
            filecolor=red,
            runcolor=red,
            urlcolor=blue,
            frenchlinks=red]{hyperref}

\begin{document}

\title{Gluon Wigner distributions in a light-cone spectator model}

\author{Chentao Tan}\affiliation{School of Physics, Southeast University, Nanjing
211189, China}

\author{Zhun Lu}
\email{zhunlu@seu.edu.cn}
\affiliation{School of Physics, Southeast University, Nanjing 211189, China}

\begin{abstract}

We study the gluon Wigner distributions of the proton which are the phase-space distributions containing the most general one-parton information.
Using the proton wave functions deduced from a light-cone spectator model that contains the gluonic degree of freedom, we calculate the Wigner distributions of the unpolarized and longitudinally polarized gluon inside the unpolarized/longitudinally polarized proton via the Fock-state overlap representation, respectively.
We present the numerical results of the transverse Wigner distributions in which the longitudinal momentum fraction is integrated out.
The mixed Wigner distributions as functions of $b_y$ and $k_x$ are also presented.
We also provide the canonical gluon orbital angular momentum and spin-orbit correlations deduced from the gluon Wigner distribution.
\end{abstract}

%\pacs{12.38.-t, 13.85.Qk, 13.88.+e}
\maketitle

\section{Introduction}\label{Sec:1}

Understanding the internal structure of nucleon is one of main quests in hadronic physics. To achieve this goal, the generalized parton distributions (GPDs)~\cite{Muller:1994ses,Goeke:2001tz,Diehl:2003ny,Ji:2004gf,Belitsky:2005qn,Boffi:2007yc} and the transverse momentum dependent parton distributions (TMDs)~\cite{Collins:1981uk,Collins:1981uw,Mulders:1995dh,Sivers:1989cc,Kotzinian:1994dv,Boer:1997nt} usually play an important role. 
The GPDs describe the parton distributions in the transverse position space through the impact parameter dependent densities ~\cite{Burkardt:2000za,Burkardt:2002hr,Bondarenko:2002pp,Riedl:2022pad}, while the TMDs describe the parton distributions in the transverse momentum space. 
Particularly, the most maximum amount of parton information is contained in a fully unintegrated, off-diagonal two-parton correlator, the so-called generalized parton correlation function (GPCF)~\cite{Meissner:2009ww,Lorce:2013pza}. 
Integrating out the parton light-cone energy from GPCFs leads to the GTMDs~\cite{Liuti:2013cna,Burkardt:2015qoa}, which are considered as the ``mother distributions" of all types of parton distributions. 
The GTMDs can not be directly measured yet at present, but there are still two major motivations to investigate them. 
On the one hand, integrating out the transverse momentum will reduce the GTMDs to the GPDs, and fixing the momentum transfer to the hadron to zero will reduce the GTMDs to the TMDs. Furthermore, for a vanishing longitudinal momentum transfer, the Fourier transform of a GTMD corresponds to a Wigner distribution. 
On the other hand, the GTMDs can be linked to the properties of the parton angular momentum. 
For example, the GTMD $F_{1,4}$ encode the canonical orbital angular momentum (OAM)~\cite{Lorce:2011kd,Hatta:2011ku,Ji:2012sj,Lorce:2012ce}, while $G_{1,1}$ contains the correlation between the longitudinal spin and the OAM of quarks~\cite{Lorce:2014mxa,Tan:2021osk}, respectively.
Furthermore, a certain combination of $F_{1,1}$, ${F_{1,2}}$ and $F_{1,3}$ encodes the total angular momentum of quarks.

The Wigner distribution~\cite{Wigner:1932eb} was initially constructed as a quantum mechanical analogue of the classical phase-space density operator. 
In QCD, the concept of the Wigner distributions was first proposed as a six-dimensional function using the nonrelativistic approximation~\cite{Ji:2003ak,Belitsky:2003nz}. 
Then the authors in Refs.~\cite{Lorce:2011dv,Lorce:2015sqe} define a five-dimensional Wigner distribution containing a longitudinal momentum fraction, two transverse momenta and two transverse positions in the light-cone framework. 
Due to the uncertainty principle, the Wigner distributions are not always positive definite over the entire phase space and have no probabilistic interpretation. 
However, a mixed distribution with probabilistic interpretation can be defined by integrating out a transverse position and a transverse momentum perpendicular to it. 
A remarkable advantage of the Wigner distribution is that it can be used to calculate the expectation value of any single-particle physical observable from its phase-space average, such as the partonic charge number~\cite{Lorce:2011kd} and OAM. 
In view of these, although the Wigner distributions are not directly observable, the model calculations are still significant for revealing the structure information of hadrons. 
The quark Wigner distributions have been calculated in various models, such as the light-cone constituent quark model~\cite{Lorce:2011ni}, the light-front dressed quark model~\cite{Mukherjee:2014nya,Mukkherjee:2015phf}, the light-cone spectator model~\cite{Liu:2014vwa,Liu:2015eqa}, the spectator diquark model~\cite{Miller:2014vla,Muller:2014tqa} and the chiral quark soliton model~\cite{Lorce:2011dv}. 
However, the gluon Wigner distributions have not been studied sufficiently, because it requires gluonic degree of freedom from the phenomenological models.
Nevertheless, a light-front dressed quark model~\cite{Mukherjee:2015aja,More:2017zqp} has been used to calculate the gluon Wigner distributions and gluon GTMDs.

In this paper, we study the Wigner distributions and the GTMDs of the unpolarized/longitudinally polarized gluon in the unpolarized/longitudinally polarized proton in a light-cone spectator model. This model has been used to calculate the twist-2 gluon TMDs~\cite{Lu:2016vqu,Bacchetta:2020vty} and GPDs~\cite{Tan:2023kbl} of gluons. 
Here, the proton target replaces the quark target, and the coupling of the nucleon-gluon-spectator vertex contains the nonperturbative effect. 
Therefore, this model not only generates the gluonic degrees of freedom, but also is closer to the realistic situation of the target state. 
We consider the minimum Fock state for the proton containing a gluon, which is a two-body composite system composed of an active gluon and a spectator particle. 
The target state is expanded in the Fock space to light-cone wave functions. 
Thus, both the wigner distributions and the GTMDs can be expressed in terms of these wave functions within the overlap representation.

The rest of the paper is organized as follows. 
In Sec.~\ref{Sec:2}, we provide the definitions of the gluon Wigner distributions and GTMDs in the proton. 
In Sec.~\ref{Sec:3}, we calculate the Wigner distributions and the GTMDs using the spectator model. 
In Sec.~\ref{Sec:4}, we present the numerical results of the Wigner distributions in the transverse position space, transverse momentum space and mixed space. 
We also discuss the canonical gluon OAM and spin-orbit correlations. In Sec.~\ref{Sec:5}, the conclusion is given.

\section{Definitions of gluon Wigner distributions and GTMDs}\label{Sec:2}

The gluon Wigner distribution in the proton is defined as~\cite{Mukherjee:2015aja,More:2017zqp,Ji:2012ba}
\begin{align}
xW^{\alpha}_{\Lambda^\prime,\Lambda}(x,\bm{k}_\perp,\bm{b}_\perp),=&\int\frac{d^2 \bm{\Delta}_\perp}{2(2\pi)^2}e^{-i\bm{\Delta}_\perp \cdot \bm{b}_\perp} \mathcal{W}^\alpha_{\Lambda^\prime,\Lambda}(x,\bm{k}_\perp,\bm{\Delta}_\perp)
\nonumber \\
=&\int \frac{d^2 \bm{\Delta}_\perp}{(2\pi)^2} e^{-i\bm{\Delta}_\perp \cdot \bm{b}_\perp} \int \frac{dz^-d^2\bm{z}_\perp}{2(2\pi)^3p^+}e^{ik\cdot z} \left\langle p^+,\frac{\bm{\Delta}_\perp}{2},\Lambda^\prime \bigg| \Gamma^{ij}F^{+i}(-\frac{z}{2}) F^{+j}(\frac{z}{2}) \bigg| p^+,-\frac{\bm{\Delta}_\perp}{2},\Lambda \right\rangle \bigg|_{z^+=0}
\label{eq:gluon wigner}
\end{align}
where $\Lambda$ ($\Lambda^\prime$) is the helicity of the initial (final) proton, $P=(p+p^\prime)/2$ is the average four-momentum of the proton, $\bm{\Delta}_\perp$ is the transverse component of the momentum transfer of the proton, $\bm{b}_\perp$ is a two-dimensional vector in the impact parameter space conjugate to $\bm{\Delta}_\perp$, and $\bm{k}_\perp$ and $x=k^+/P^+$ are the transverse momentum and the average light-front momentum fraction carried by the active gluon, respectively. 
For different polarizations of gluons, $\alpha=1,2,3,4$ represent unpolarization (U), longitudinal polarization (L) and two linear polarizations ($\mathcal{T}$), respectively. The gluon field strength tensor is given by
\begin{align}
F_a^{+i}=\partial^+ A_a^i-\partial^i A_a^+ + g f_{abc} A_b^+ A_c^i.
\end{align}
$\mathcal{W}^\alpha_{\Lambda^\prime,\Lambda}$ in Eq.~(\ref{eq:gluon wigner}) denotes the generalized gluon correlator at zero skewness ($\xi=0$), and it can be parameterized to a complete list of gluon GTMDs for a spin-1/2 target~\cite{Lorce:2013pza}. 
This correlator needs two gauge links to ensure the color
gauge invariance. We choose the light-cone gauge $A^+=0$ and take the gauge link to be unity.

At leading-twist, there are sixteen gluon GTMDs in total~\cite{Lorce:2013pza}. Here we only focus on the cases of $\alpha=1,2$, which correspond to $\Gamma^{ij}=\delta^{ij}_\perp,-i\epsilon_\perp^{ij}$:
\begin{align}
\mathcal{W}^1_{\Lambda^\prime,\Lambda}(x,\bm{k}_\perp,\bm{\Delta}_\perp)=&\frac{1}{M} \bigg\{ \left[M\delta_{\Lambda^\prime,\Lambda}-\frac{1}{2}(\Lambda\Delta_\perp^1+i\Delta_\perp^2)\delta_{-\Lambda^\prime,\Lambda}\right]F^g_{1,1} +(\Lambda k_\perp^1+ik_\perp^2)\delta_{-\Lambda^\prime,\Lambda}F_{1,2}^g \nonumber \\
&+(\Lambda\Delta_\perp^1+i\Delta_\perp^2)\delta_{-\Lambda^\prime,\Lambda}F^g_{1,3} +\frac{i\epsilon_\perp^{ij}k_\perp^i\Delta_\perp^j}{M} \Lambda \delta_{\Lambda^\prime,\Lambda}F_{1,4}^g \bigg\},
\label{eq:f GTMDs}\\
\mathcal{W}^2_{\Lambda^\prime,\Lambda}(x,\bm{k}_\perp,\bm{\Delta}_\perp)=&\frac{1}{M} \bigg\{ -\frac{i\epsilon^{ij}_\perp k_\perp^i\Delta_\perp^j}{M^2} \left[M\delta_{\Lambda^\prime,\Lambda}-\frac{1}{2}(\Lambda\Delta_\perp^1+i\Delta_\perp^2)\delta_{-\Lambda^\prime,\Lambda}\right]G^g_{1,1} +(k_\perp^1+i\Lambda k_\perp^2)\delta_{-\Lambda^\prime,\Lambda}G_{1,2}^g \nonumber \\
&+(\Delta_\perp^1+i\Lambda\Delta_\perp^2)\delta_{-\Lambda^\prime,\Lambda}G^g_{1,3} +\Lambda M \delta_{\Lambda^\prime,\Lambda}G_{1,4}^g \bigg\},
\label{eq:g GTMDs}
\end{align}
where $M$ is the proton mass, $\delta^{ij}_{\perp}=-g^{ij}_\perp$ and $\epsilon^{ij}_\perp=\epsilon^{+-ij}$ is an antisymmetric tensor with $\epsilon^{+-12}=1$. Four twist-2 F-type GTMDs in Eq.~(\ref{eq:f GTMDs}) describe the distributions of unpolarized gluons, while four twist-2 G-type GTMDs in Eq.~(\ref{eq:g GTMDs}) describe the gluon helicity distributions. 
The linearly polarized gluons are described by the other eight twist-2 H-type GTMDs are needed. 
GTMDs are functions of the set of variables $x$, $\xi$, $\bm{k}_\perp^2$, $\bm{k}_\perp \cdot \bm{\Delta}_\perp$ and $\bm{\Delta}_\perp^2$. 
In general, the GTMDs are complex-valued functions~\cite{Meissner:2008ay,Meissner:2009ww}, while the Wigner distributions are real~\cite{Lorce:2011kd}. 

Specifically, the gluon GTMDs can be expressed through the generalized correlator $\mathcal{W}^\alpha_{\Lambda^\prime,\Lambda}$ as
\begin{align}
\mathcal{W}^1_{\uparrow,\uparrow}(x,\bm{k}_\perp,\bm{\Delta}_\perp)+\mathcal{W}^1_{\uparrow,\uparrow}(x,-\bm{k}_\perp,\bm{\Delta}_\perp)&=2F_{1,1}^g
\label{eq:f11},\\
\mathcal{W}^1_{\uparrow,\uparrow}(x,\bm{k}_\perp,\bm{\Delta}_\perp)-\mathcal{W}^1_{\uparrow,\uparrow}(x,-\bm{k}_\perp,\bm{\Delta}_\perp)&=2 \frac{i\epsilon_\perp^{ij}k_\perp^i \Delta_\perp^j}{M^2} F_{1,4}^g
\label{eq:f12},\\
\mathcal{W}^1_{\downarrow,\uparrow}(x,\bm{k}_\perp,\bm{\Delta}_\perp)+\mathcal{W}^1_{\downarrow,\uparrow}(x,-\bm{k}_\perp,\bm{\Delta}_\perp)&=2\frac{1}{M}\left[-\frac{1}{2} (\Delta_\perp^1+i\Delta_\perp^2)F_{1,1}^g+(\Delta_\perp^1+i\Delta_\perp^2)F_{1,3}^g\right]
\label{eq:f13},\\
\mathcal{W}^1_{\downarrow,\uparrow}(x,\bm{k}_\perp,\bm{\Delta}_\perp)-\mathcal{W}^1_{\downarrow,\uparrow}(x,-\bm{k}_\perp,\bm{\Delta}_\perp)&=2\frac{1}{M} (k_\perp^1+ik_\perp^2)F_{1,2}^g
\label{eq:f14};
\end{align}
and
\begin{align}
\mathcal{W}^2_{\uparrow,\uparrow}(x,\bm{k}_\perp,\bm{\Delta}_\perp)+\mathcal{W}^2_{\uparrow,\uparrow}(x,-\bm{k}_\perp,\bm{\Delta}_\perp)&=2G_{1,4}^g
\label{eq:g11},\\
\mathcal{W}^2_{\uparrow,\uparrow}(x,\bm{k}_\perp,\bm{\Delta}_\perp)-\mathcal{W}^2_{\uparrow,\uparrow}(x,-\bm{k}_\perp,\bm{\Delta}_\perp)&=-2 \frac{i\epsilon_\perp^{ij}k_\perp^i \Delta_\perp^j}{M^2} G_{1,1}^g
\label{eq:g12},\\
\mathcal{W}^2_{\downarrow,\uparrow}(x,\bm{k}_\perp,\bm{\Delta}_\perp)+\mathcal{W}^2_{\downarrow,\uparrow}(x,-\bm{k}_\perp,\bm{\Delta}_\perp)&=2\frac{1}{M} (\Delta_\perp^1+i\Delta_\perp^2)G_{1,3}^g
\label{eq:g13},\\
\mathcal{W}^2_{\downarrow,\uparrow}(x,\bm{k}_\perp,\bm{\Delta}_\perp)-\mathcal{W}^2_{\downarrow,\uparrow}(x,-\bm{k}_\perp,\bm{\Delta}_\perp)&=2\frac{1}{M} \left[ \frac{i\epsilon_\perp^{ij} k_\perp^i \Delta_\perp^j}{2M^2} (\Delta_\perp^1+i\Delta_\perp^2)G_{1,1}^g+(k_\perp^1+ik_\perp^2)G_{1,2}^g \right]
\label{eq:g14},
\end{align}
where $\uparrow$ ($\downarrow$) denotes that the helicity of the proton is $+1/2$ $(-1/2)$.

The gluon and the proton are either unpolarized or polarized along three orthogonal directions, which means that there are also sixteen Wigner distributions at leading twist. These Wigner distributions can be written in terms of sixteen independent linear combinations of the leading-twist GTMDs. In the following, we consider the gluon Wigner distributions without any transverse polarization~\cite{Mukherjee:2015aja}:
\begin{align}
W_{UU}(x,\bm{k}_\perp,\bm{b}_\perp)&=W^1_{\uparrow,\uparrow}(x,\bm{k}_\perp,\bm{b}_\perp)+W^1_{\downarrow,\downarrow}(x,\bm{k}_\perp,\bm{b}_\perp)
\end{align}
represents the Wigner distribution of the unpolarized gluon in the unpolarized target;
\begin{align}
W_{LU}(x,\bm{k}_\perp,\bm{b}_\perp)&=W^1_{\uparrow,\uparrow}(x,\bm{k}_\perp,\bm{b}_\perp)-W^1_{\downarrow,\downarrow}(x,\bm{k}_\perp,\bm{b}_\perp)
\end{align}
represents the Wigner distribution of the unpolarized gluon in the longitudinally polarized target;
\begin{align}
W_{UL}(x,\bm{k}_\perp,\bm{b}_\perp)&=W^2_{\uparrow,\uparrow}(x,\bm{k}_\perp,\bm{b}_\perp)+W^2_{\downarrow,\downarrow}(x,\bm{k}_\perp,\bm{b}_\perp)
\end{align}
represents the Wigner distribution of the longitudinally polarized gluon in the unpolarized target; and
\begin{align}
W_{LL}(x,\bm{k}_\perp,\bm{b}_\perp)&=W^2_{\uparrow,\uparrow}(x,\bm{k}_\perp,\bm{b}_\perp)-W^2_{\downarrow,\downarrow}(x,\bm{k}_\perp,\bm{b}_\perp)
\end{align}
represents the Wigner distribution of the longitudinally polarized gluon in the longitudinally polarized target,
where $W^\alpha_{\Lambda^\prime,\Lambda}(x,\bm{k}_\perp,\bm{b}_\perp)$ comes from the Eq.~(\ref{eq:gluon wigner}).

In addition, some relations between the Wigner distributions and the GTMDs are given by~\cite{Lorce:2011kd}
\begin{align}
W_{UU}(x,\bm{k}_\perp,\bm{b}_\perp)&=\mathcal{F}_{1,1}^g(x,0,\bm{k}^2_\perp,\bm{k}_\perp \cdot \bm{b}_\perp,\bm{b}^2_\perp),\\
W_{LU}(x,\bm{k}_\perp,\bm{b}_\perp)&=-\frac{1}{M^2}\epsilon_\perp^{ij} k_\perp^i \frac{\partial}{\partial b_\perp^j} \mathcal{F}_{1,4}^g(x,0,\bm{k}^2_\perp,\bm{k}_\perp \cdot \bm{b}_\perp,\bm{b}^2_\perp)
\label{eq:wlu f14},\\
W_{UL}(x,\bm{k}_\perp,\bm{b}_\perp)&=\frac{1}{M^2}\epsilon_\perp^{ij} k_\perp^i \frac{\partial}{\partial b_\perp^j} \mathcal{G}_{1,1}^g(x,0,\bm{k}^2_\perp,\bm{k}_\perp \cdot \bm{b}_\perp,\bm{b}^2_\perp)
\label{eq:wul g11},\\
W_{LL}(x,\bm{k}_\perp,\bm{b}_\perp)&=\mathcal{G}_{1,4}^g(x,0,\bm{k}^2_\perp,\bm{k}_\perp \cdot \bm{b}_\perp,\bm{b}^2_\perp),
\end{align}
where $\mathcal{X}$ are the Fourier transform of the corresponding GTMD $X$
\begin{align}
\mathcal{X}(x,0,\bm{k}^2_\perp,\bm{k}_\perp \cdot \bm{b}_\perp,\bm{b}^2_\perp)=\int\frac{d^2\bm{\Delta}_\perp}{(2\pi)^2}e^{-i\bm{\Delta}_\perp \cdot \bm{b}_\perp} X(x,0,\bm{k}^2_\perp,\bm{k}_\perp \cdot \bm{\Delta}_\perp,\bm{\Delta}^2_\perp).
\end{align}

\begin{align}
\int d^2 \bm b_\perp \mathcal{X}(x,0,\bm{k}^2_\perp,\bm{k}_\perp \cdot \bm{b}_\perp,\bm{b}^2_\perp)=&\int d^2 \bm b_\perp\int\frac{d^2\bm{\Delta}_\perp}{(2\pi)^2}e^{-i\bm{\Delta}_\perp \cdot \bm{b}_\perp} X(x,0,\bm{k}^2_\perp,\bm{k}_\perp \cdot \bm{\Delta}_\perp,\bm{\Delta}^2_\perp)\\
=& \int {d^2\bm{\Delta}_\perp}\delta^2(\bm{\Delta}_\perp) X(x,0,\bm{k}^2_\perp,\bm{k}_\perp \cdot \bm{\Delta}_\perp,\bm{\Delta}^2_\perp)\\
=& X(x,0,\bm{k}^2_\perp,0,0)
\end{align}

\section{Wigner distributions and GTMDs in the overlap representation}\label{Sec:3}

In this section, we express the generalized gluon correlators in the overlap representation within a light-cone spectator model and calculate the corresponding gluon Wigner distributions and GTMDs. 
The light-cone formalism~\cite{Brodsky:1997de} has been widely used to calculate the form factors~\cite{Lepage:1979za}, PDFs~\cite{Bacchetta:2008af} and the GPDs~\cite{Brodsky:2000xy} of nucleons and mesons. 
On the other hand, the overlap representation has also been applied to study various form factors of hadrons~\cite{Brodsky:2000ii,Lu:2006kt,Xiao:2003wf}. 
The gluon Wigner distributions in the nucleon have been calculated in the light-front dressed quark model~\cite{Mukherjee:2015aja,More:2017zqp}. 
Therefore, it is natural to extend this model to a model containing the proton target. The minimum Fock state for a proton containing a gluon is $|qqqg\rangle$. Due to the complexity of the four-body system, we treat this target state as a two-body composite system composed of an active gluon $g$ and a spectator particle $X$~~\cite{Lu:2016vqu,Bacchetta:2020vty}:
\begin{align}
|p;S \rangle \rightarrow |g^{s_g}X^{s_X}(uud)\rangle ,
\end{align}
where $s_g$ and $s_X$ denote the spins of the gluon and the spectator particle, respectively.
In principle, the spin quantum number of the spectator can be $s_{\rm{X}}=1/2$ or $3/2$.
According to the angular momentum conservation, for a $s_X={1\over 2}$ spectator, the orbital angular momentum of the gluon may be $L=0$ or $L=1$. While for a $s_X={3\over 2}$ spectator, the orbital angular momentum of the gluon has to be at least $L=2$. 
As the contribution from high OAM components is assumed to be much smaller than that from the low OAM components, in this calculation we only consider the spin-1/2 component and ignore the contribution from a $s_X={3\over 2}$ spectator, following Refs.~\cite{Lu:2016vqu,Bacchetta:2020vty}.

For a proton with $J_z=+1/2$, the Fock-state expansion is given by
\begin{align}
|\Psi^\uparrow_{\text{two\, particle}}(p^+,\bm{p}_\perp=\bm{0}_\perp) \rangle =&\int \frac{d^2\bm{k}_\perp dx}{16\pi^3\sqrt{x(1-x)}}\nonumber \\
&\times \left[\psi^\uparrow_{+1+\frac{1}{2}}(x,\bm{k}_\perp)\left|+1,+\frac{1}{2};xp^+,\bm{k}_\perp \right\rangle+\psi^\uparrow_{+1-\frac{1}{2}}(x,\bm{k}_\perp)\left|+1,-\frac{1}{2};xp^+,\bm{k}_\perp \right\rangle \nonumber \right.\\
&\left.+\psi^\uparrow_{-1+\frac{1}{2}}(x,\bm{k}_\perp)\left|-1,+\frac{1}{2};xp^+,\bm{k}_\perp \right\rangle+\psi^\uparrow_{-1-\frac{1}{2}}(x,\bm{k}_\perp)\left|-1,-\frac{1}{2};xp^+,\bm{k}_\perp \right\rangle \right],
\label{eq:fockstate}
\end{align}
where $\psi^\uparrow_{s_g^z s_X^z}(x,\bm{k}_\perp)$ are the light-cone wave functions corresponding to the two-particle states $|s_g^z,s_X^z;xP^+,\bm{k}_\perp \rangle$, with $s_g^z$ and $s_X^z$ denoting the $z$ components of the gluon and spectator spins, respectively. These wave functions have the forms
\begin{align}
\psi^\uparrow_{+1+\frac{1}{2}}(x,\bm{k}_\perp)&=-\sqrt{2}\frac{-k^1_\perp+ik^2_\perp}{x(1-x)}\phi,\nonumber\\
\psi^\uparrow_{+1-\frac{1}{2}}(x,\bm{k}_\perp)&=-\sqrt{2}\left(M-\frac{M_X}{1-x}\right)\phi,\nonumber\\
\psi^\uparrow_{-1+\frac{1}{2}}(x,\bm{k}_\perp)&=-\sqrt{2}\frac{+k^1_\perp+ik^2_\perp}{x}\phi,\nonumber\\
\psi^\uparrow_{-1-\frac{1}{2}}(x,\bm{k}_\perp)&=0,
\label{eq:wavefunction+}
\end{align}
where
\begin{align}
\phi(x,\bm{k}_\perp)=\frac{\lambda \sqrt{x} (1-x)}{x(1-x)M^2-(1-x)(\bm{k}^2_\perp+M_g^2)-x(\bm{k}_\perp^2+M_X^2)},
\end{align}
is the wave function in the momentum space, $M_X$ is the spectator mass, $M_g$ is the gluon mass, and $\lambda$ is the coupling of the nucleon-gluon-spectator vertex. In order to simulate the nonperturbative effect of the vertex, we choose the Brodsky-Hwang-Lepage prescription for the coupling $\lambda$~\cite{Brodsky:1982nx}:
\begin{align}
\lambda \rightarrow N_\lambda \text{exp}(-\frac{\mathcal{M}^2}{2\beta^2_1}),
\end{align}
where $N_\lambda$ is a strength parameter for the vertex, $\beta_1$ is a cutting-off parameter, and $\mathcal{M}$ is the invariant mass of the two-particle system
\begin{align}
\mathcal{M}^2=\frac{\bm{k}^2_\perp+M_g^2}{x}+\frac{\bm{k}^2_\perp+M_X^2}{1-x}.
\end{align}

Similarly, for a proton with $J_z=-1/2$, the Fock-state expansion has the form
\begin{align}
|\Psi^\downarrow_{\text{two\, particle}}(p^+,\bm{p}_\perp=\bm{0}_\perp) \rangle =&\int \frac{d^2\bm{k}_\perp dx}{16\pi^3\sqrt{x(1-x)}}\nonumber \\
&\times \left[\psi^\downarrow_{+1+\frac{1}{2}}(x,\bm{k}_\perp)\left|+1,+\frac{1}{2};xp^+,\bm{k}_\perp \right\rangle+\psi^\downarrow_{+1-\frac{1}{2}}(x,\bm{k}_\perp)\left|+1,-\frac{1}{2};xp^+,\bm{k}_\perp \right\rangle \nonumber \right.\\
&\left.+\psi^\downarrow_{-1+\frac{1}{2}}(x,\bm{k}_\perp)\left|-1,+\frac{1}{2};xp^+,\bm{k}_\perp \right\rangle+\psi^\downarrow_{-1-\frac{1}{2}}(x,\bm{k}_\perp)\left|-1,-\frac{1}{2};xp^+,\bm{k}_\perp \right\rangle \right],
\end{align}
where
\begin{align}
\psi^\downarrow_{+1+\frac{1}{2}}(x,\bm{k}_\perp)&=0,\nonumber\\
\psi^\downarrow_{+1-\frac{1}{2}}(x,\bm{k}_\perp)&=-\sqrt{2}\frac{-k^1_\perp+ik^2_\perp}{x}\phi,\nonumber\\
\psi^\downarrow_{-1+\frac{1}{2}}(x,\bm{k}_\perp)&=-\sqrt{2}\left(M-\frac{M_X}{1-x}\right)\phi,\nonumber\\
\psi^\downarrow_{-1-\frac{1}{2}}(x,\bm{k}_\perp)&=-\sqrt{2}\frac{+k^1_\perp+ik^2_\perp}{x(1-x)}\phi.
\label{eq:wavefunction-}
\end{align}

Based on the above results, the generalized gluon correlators in the overlap representation can be expressed as
\begin{align}
\mathcal{W}^1_{\Lambda,\Lambda}(x,\bm{k}_\perp,\bm{\Delta}_\perp)&=N\sum_{s_{g}^{z},s_X^z}\psi^{\Lambda\star}_{s_{g}^z,s_X^g}(x^{out},\bm{k}_\perp^{out}) \psi^{\Lambda}_{s_{g}^z,s_X^g}(x^{in},\bm{k}_\perp^{in})(\epsilon_{s_{g}^z}^1\epsilon_{s_{g}^z}^{1\ast}+\epsilon_{s_{g}^z}^2\epsilon_{s_{g}^z}^{2\ast})
\label{eq:w1+},\\
\mathcal{W}^1_{\downarrow,\uparrow}(x,\bm{k}_\perp,\bm{\Delta}_\perp)&=-N\sum_{s_{g}^{z},s_X^z}\psi^{\downarrow\star}_{s_{g}^z,s_X^g}(x^{out},\bm{k}_\perp^{out}) \psi^{\uparrow}_{s_{g}^z,s_X^g}(x^{in},\bm{k}_\perp^{in})(\epsilon_{s_{g}^z}^1\epsilon_{s_{g}^z}^{1\ast}+\epsilon_{s_{g}^z}^2\epsilon_{s_{g}^z}^{2\ast})
\label{eq:w1-},\\
\mathcal{W}^2_{\Lambda,\Lambda}(x,\bm{k}_\perp,\bm{\Delta}_\perp)&=iN\sum_{s_{g}^{z},s_X^z}\psi^{\Lambda\star}_{s_{g}^z,s_X^g}(x^{out},\bm{k}_\perp^{out}) \psi^{\Lambda}_{s_{g}^z,s_X^g}(x^{in},\bm{k}_\perp^{in})(\epsilon_{s_{g}^z}^1\epsilon_{s_{g}^z}^{2\ast}-\epsilon_{s_{g}^z}^2\epsilon_{s_{g}^z}^{1\ast})
\label{eq:w2+},\\
\mathcal{W}^2_{\downarrow,\uparrow}(x,\bm{k}_\perp,\bm{\Delta}_\perp)&=-iN\sum_{s_{g}^{z},s_X^z}\psi^{\downarrow\star}_{s_{g}^z,s_X^g}(x^{out},\bm{k}_\perp^{out}) \psi^{\uparrow}_{s_{g}^z,s_X^g}(x^{in},\bm{k}_\perp^{in})(\epsilon_{s_{g}^z}^1\epsilon_{s_{g}^z}^{2\ast}-\epsilon_{s_{g}^z}^2\epsilon_{s_{g}^z}^{1\ast})
\label{eq:w2-},
\end{align}
where $N=1/(2(2\pi)^3)$. The gluon polarization vectors $\epsilon_{s_g^z}^\mu$ read
\begin{align}
\epsilon^\mu_{\pm}=(0,0,\bm{\epsilon}_{\pm})=\frac{1}{\sqrt{2}}(0,0,\mp1,-i).
\end{align}
The arguments of the initial-state and final-state wave functions are given by
\begin{align}
\bm{k}^{\text{in}}_\perp=&\bm{k}_\perp - (1-x^{\text{in}})\frac{\bm{\Delta}_\perp}{2},\qquad \text{with} \qquad x^{\text{in}}=\frac{x+\xi}{1+\xi}, \\
\bm{k}^{\text{out}}_\perp=&\bm{k}_\perp +(1-x^{\text{out}}) \frac{\bm{\Delta}_\perp}{2},\qquad \text{with} \qquad x^{\text{out}}=\frac{x-\xi}{1-\xi},
\end{align}
respectively.

Combining the light-cone wave functions in Eqs.~(\ref{eq:wavefunction+},\ref{eq:wavefunction-}) and the overlap representation for $\mathcal{W}_{\Lambda^\prime,\Lambda}^\alpha$ in Eqs.~(\ref{eq:w1+}-\ref{eq:w2-}), the gluon Wigner distributions in our model have the forms
\begin{align}
W_{UU}(x,\bm{k}_\perp,\bm{b}_\perp)=&NN_\lambda^2\int\frac{d^2 \bm{\Delta}_\perp}{2(2\pi)^2}e^{-i\bm{\Delta}_\perp \cdot \bm{b}_\perp} \text{exp}\left({-\frac{4\bm{k}_\perp^2+4x M_X^2+(1-x)^2\bm{\Delta}^2_\perp}{4x(1-x)\beta_1^2}}\right) \nonumber \\
&\frac{4x^2(M_X-M(1-x))^2+(1+(1-x)^2)(4\bm{k}_\perp^2-(1-x)^2\bm{\Delta}_\perp^2)}{xD_{LSM}^g(x,\bm{k}_\perp,\bm{\Delta}_\perp)},\\
W_{LU}(x,\bm{k}_\perp,\bm{b}_\perp)=&NN_\lambda^2\int\frac{d^2 \bm{\Delta}_\perp}{2(2\pi)^2}e^{-i\bm{\Delta}_\perp \cdot \bm{b}_\perp} \text{exp}\left({-\frac{4\bm{k}_\perp^2+4x M_X^2+(1-x)^2\bm{\Delta}^2_\perp}{4x(1-x)\beta_1^2}}\right) \nonumber \\
&\frac{4i(1-x)(2-x)(k_x\Delta_y-k_y\Delta_x)}{D_{LSM}^g(x,\bm{k}_\perp,\bm{\Delta}_\perp)},\\
W_{UL}(x,\bm{k}_\perp,\bm{b}_\perp)=&NN_\lambda^2\int\frac{d^2 \bm{\Delta}_\perp}{2(2\pi)^2}e^{-i\bm{\Delta}_\perp \cdot \bm{b}_\perp} \text{exp}\left({-\frac{4\bm{k}_\perp^2+4x M_X^2+(1-x)^2\bm{\Delta}^2_\perp}{4x(1-x)\beta_1^2}}\right) \nonumber \\
&\frac{4i(1-x)(1+(1-x)^2)(k_x\Delta_y-k_y\Delta_x)}{xD_{LSM}^g(x,\bm{k}_\perp,\bm{\Delta}_\perp)},\\
W_{LL}(x,\bm{k}_\perp,\bm{b}_\perp)=&NN_\lambda^2\int\frac{d^2 \bm{\Delta}_\perp}{2(2\pi)^2}e^{-i\bm{\Delta}_\perp \cdot \bm{b}_\perp} \text{exp}\left({-\frac{4\bm{k}_\perp^2+4x M_X^2+(1-x)^2\bm{\Delta}^2_\perp}{4x(1-x)\beta_1^2}}\right) \nonumber \\
&\frac{4x(M_X-M(1-x))^2+(2-x)(4\bm{k}_\perp^2-(1-x)^2\bm{\Delta}_\perp^2)}{D_{LSM}^g(x,\bm{k}_\perp,\bm{\Delta}_\perp)},
\end{align}
where
\begin{align}
D_{LSM}^g(x,\bm{k}_\perp,\bm{\Delta}_\perp)=[(\bm{k}_\perp-\frac{1}{2}(1-x)\bm{\Delta}_\perp)^2+xM_X^2-x(1-x)M^2][(\bm{k}_\perp+\frac{1}{2}(1-x)\bm{\Delta}_\perp)^2+xM_X^2-x(1-x)M^2].
\end{align}

Then the gluon GTMDs are given by
\begin{align}
F_{1,1}^g=&\frac{NN_\lambda^2}{2} \text{exp}\left({-\frac{4\bm{k}_\perp^2+4x M_X^2+(1-x)^2\bm{\Delta}^2_\perp}{4x(1-x)\beta_1^2}}\right) \frac{4x^2(M_X-M(1-x))^2+(1+(1-x)^2)(4\bm{k}_\perp^2-(1-x)^2\bm{\Delta}_\perp^2)}{xD_{LSM}^g(x,\bm{k}_\perp,\bm{\Delta}_\perp)},\\
F_{1,2}^g=&0,\\
F_{1,3}^g=&\frac{1}{2}F_{1,1}^g-\frac{NN_\lambda^2}{2} \text{exp}\left({-\frac{4\bm{k}_\perp^2+4x M_X^2+(1-x)^2\bm{\Delta}^2_\perp}{4x(1-x)\beta_1^2}}\right) \frac{4(M_X-(1-x)M)M(1-x)^2}{D_{LSM}^g(x,\bm{k}_\perp,\bm{\Delta}_\perp)},\\
F_{1,4}^g=&\frac{NN_\lambda^2}{2} \text{exp}\left({-\frac{4\bm{k}_\perp^2+4x M_X^2+(1-x)^2\bm{\Delta}^2_\perp}{4x(1-x)\beta_1^2}}\right)\frac{4(1-x)(2-x)M^2}{D_{LSM}^g(x,\bm{k}_\perp,\bm{\Delta}_\perp)},
\end{align}
and
\begin{align}
G_{1,1}^g=&-\frac{NN_\lambda^2}{2} \text{exp}\left({-\frac{4\bm{k}_\perp^2+4x M_X^2+(1-x)^2\bm{\Delta}^2_\perp}{4x(1-x)\beta_1^2}}\right) \frac{4(1-x)(1+(1-x)^2)M^2}{xD_{LSM}^g(x,\bm{k}_\perp,\bm{\Delta}_\perp)},\\
G_{1,2}^g=&\frac{(k_x\Delta_y-k_y\Delta_x)^2}{2M^2\bm{k}_\perp^2}G_{1,1}^g-\frac{NN_\lambda^2}{2} \text{exp}\left({-\frac{4\bm{k}_\perp^2+4x M_X^2+(1-x)^2\bm{\Delta}^2_\perp}{4x(1-x)\beta_1^2}}\right) \frac{8(M_X-(1-x)M)M(1-x)}{D_{LSM}^g(x,\bm{k}_\perp,\bm{\Delta}_\perp)},\\
G_{1,3}^g=&0,\\
G_{1,4}^g=&\frac{NN_\lambda^2}{2} \text{exp}\left({-\frac{4\bm{k}_\perp^2+4x M_X^2+(1-x)^2\bm{\Delta}^2_\perp}{4x(1-x)\beta_1^2}}\right)
\frac{4x(M_X-M(1-x))^2+(2-x)(4\bm{k}_\perp^2-(1-x)^2\bm{\Delta}_\perp^2)}{D_{LSM}^g(x,\bm{k}_\perp,\bm{\Delta}_\perp)}.
\end{align}

At the TMD-limit and GPD-limit, the GTMD $F_{1,1}^g$ reduce to $f_1^g$ and $H^g$, respectively
\begin{align}
f_1^g(x,\bm{k}_\perp^2)=&F_{1,1}^g(x,0,\bm{k}^2_\perp,0,0),\\
H^g(x,0,\bm{\Delta}^2_\perp)=&\int d^2\bm{k}_\perp F_{1,1}^g(x,0,\bm{k}^2_\perp,\bm{k}_\perp \cdot \bm{\Delta}_\perp,\bm{\Delta}^2_\perp);
\end{align}
and $G_{1,4}^g$ reduces to $g_{1L}^g$ and $\tilde{H}^g$
\begin{align}
g_{1L}^g(x,\bm{k}_\perp^2)=&G_{1,4}^g(x,0,\bm{k}^2_\perp,0,0),\\
\tilde{H}^g(x,0,\bm{\Delta}^2_\perp)=&\int d^2\bm{k}_\perp G_{1,4}^g(x,0,\bm{k}^2_\perp,\bm{k}_\perp \cdot \bm{\Delta}_\perp,\bm{\Delta}^2_\perp).
\end{align}
Moreover, the GPD $E^g$ is defined as
\begin{align}
E^g(x,0,\bm{\Delta}^2_\perp)=&\int d^2\bm{k}_\perp \bigg[-F_{1,1}^g(x,0,\bm{k}^2_\perp,\bm{k}_\perp \cdot \bm{\Delta}_\perp,\bm{\Delta}^2_\perp)+2\bigg(\frac{\bm{k}_\perp \cdot \bm{\Delta}_\perp}{\bm{\Delta}^2_\perp}F_{1,2}^g(x,0,\bm{k}^2_\perp,\bm{k}_\perp \cdot \bm{\Delta}_\perp,\bm{\Delta}^2_\perp) \nonumber \\
&+F_{1,3}^g(x,0,\bm{k}^2_\perp,\bm{k}_\perp \cdot \bm{\Delta}_\perp,\bm{\Delta}^2_\perp)\bigg)\bigg].
\end{align}
As a check, one can find that using the GTMDs derived in this section, the results in Refs.~\cite{Kaur:2019kpe,Tan:2023kbl} can be obtained.

\section{Numerical results}\label{Sec:4}

In order to present the numerical results of the gluon Wigner distributions in the proton, we need to specify the values of the parameters $N_\lambda$, $M_X$, $\beta_1$ and $M$ in the model~\cite{Lu:2016vqu}:
\begin{align}
N_\lambda=5.026, \quad M_X=0.943\ GeV ,\nonumber \\
\beta_1=2.092\ GeV, \quad M=0.938\ GeV
\label{eq:parameter}.
\end{align}

The Wigner distribution is a five-dimensional phase space distribution, which contains a longitudinal momentum fraction ($x$) and two transverse components ($\bm{k}_\perp,\bm{b}_\perp$). 
In the following, we consider the first $x$-moment of the Wigner distributions
\begin{align}
W(\bm{k}_\perp,\bm{b}_\perp)=\int^1_0 dx W(x,\bm{k}_\perp,\bm{b}_\perp),
\end{align}
which are referred to as the transverse Wigner distributions.

\begin{figure}
  \centering
  \includegraphics[width=0.42\columnwidth]{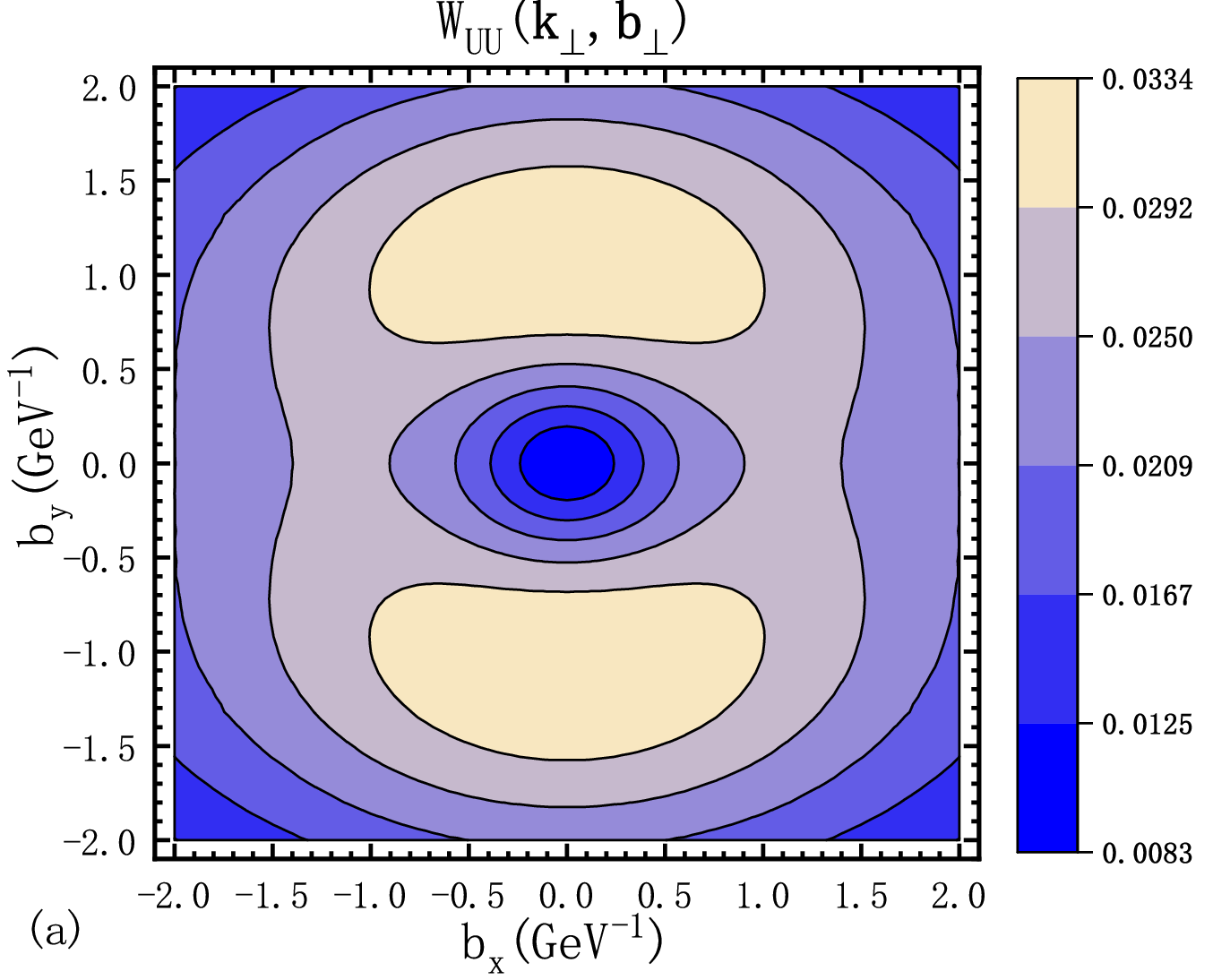}~~~
  \includegraphics[width=0.42\columnwidth]{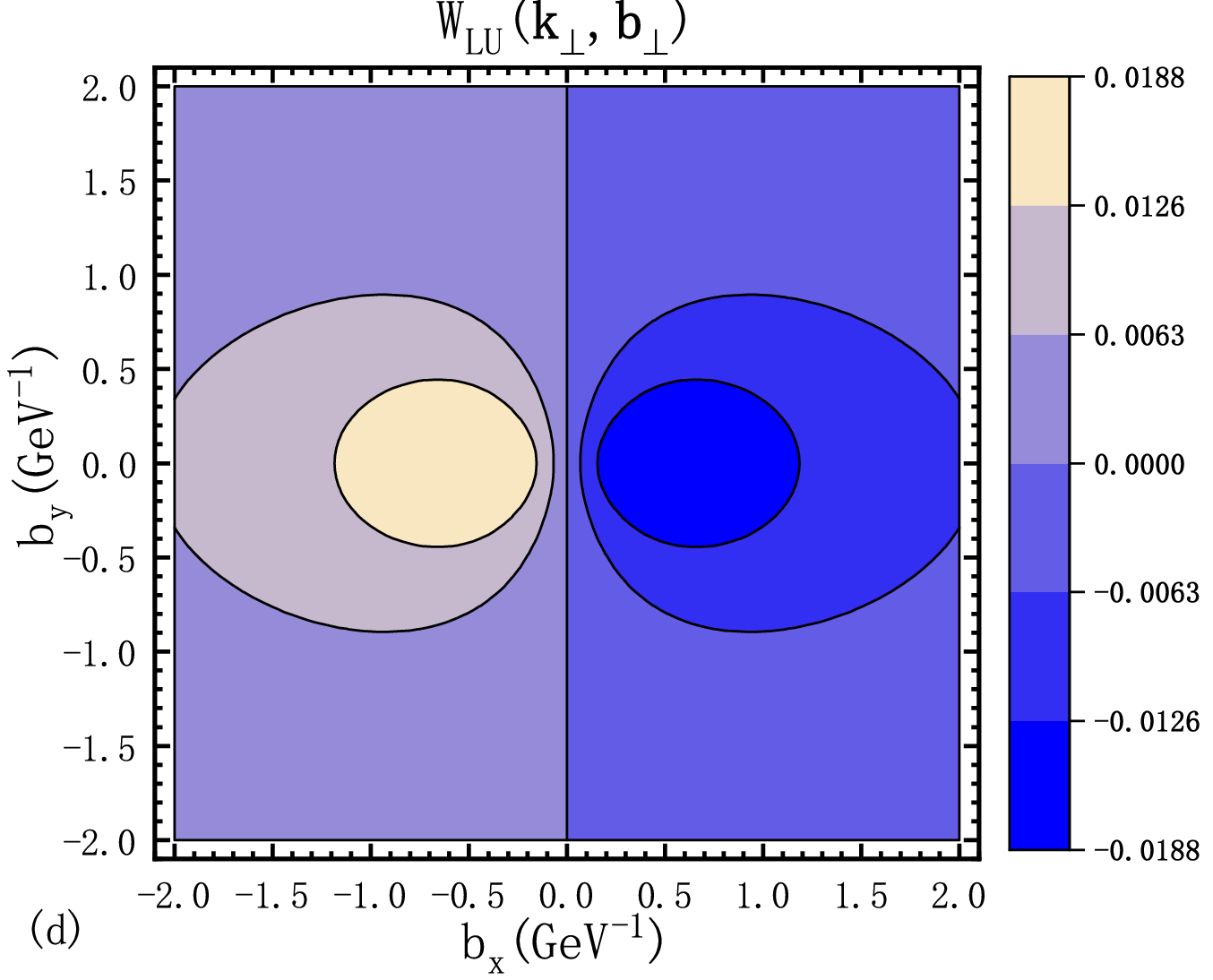}\\
  \includegraphics[width=0.42\columnwidth]{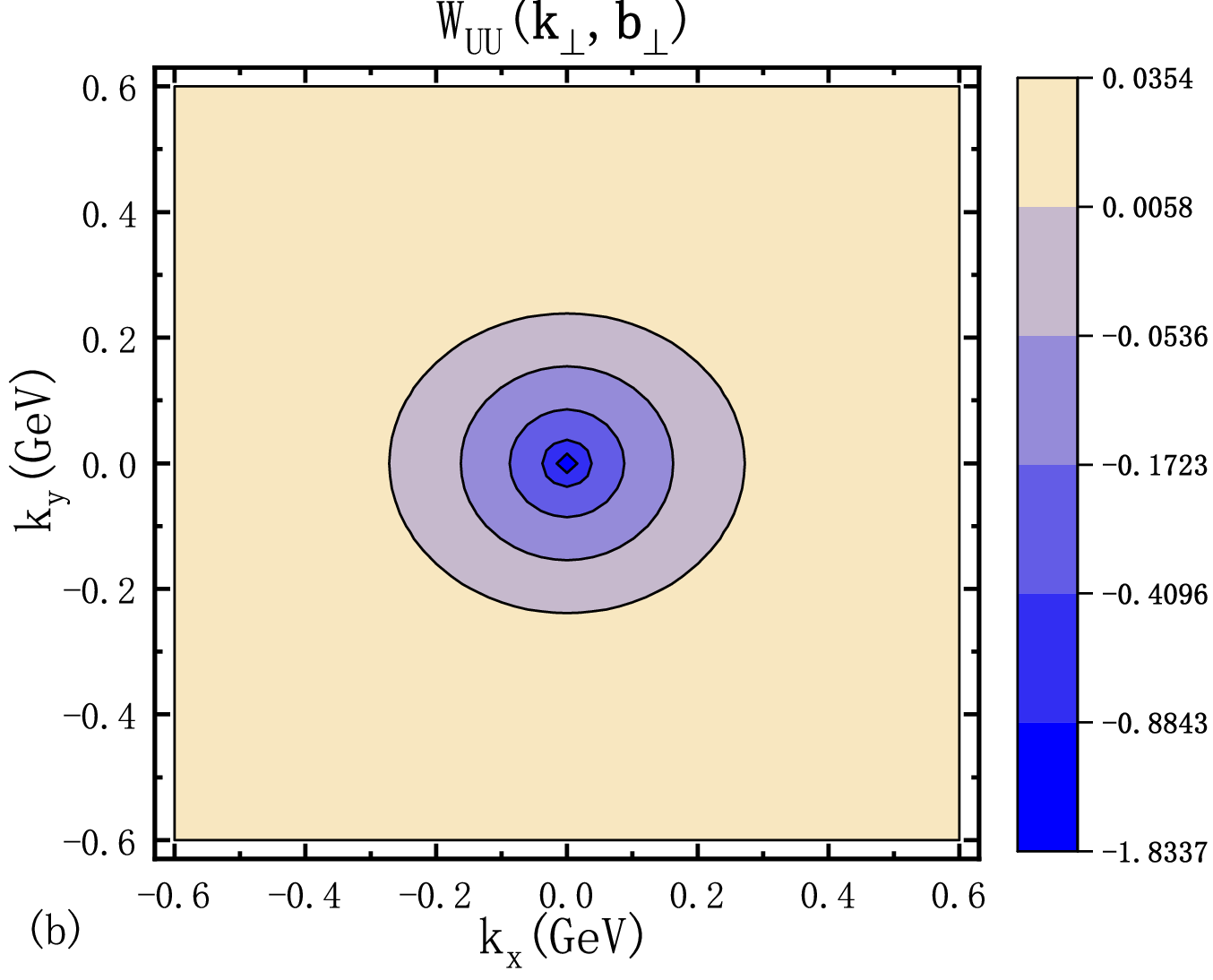}~~~
  \includegraphics[width=0.42\columnwidth]{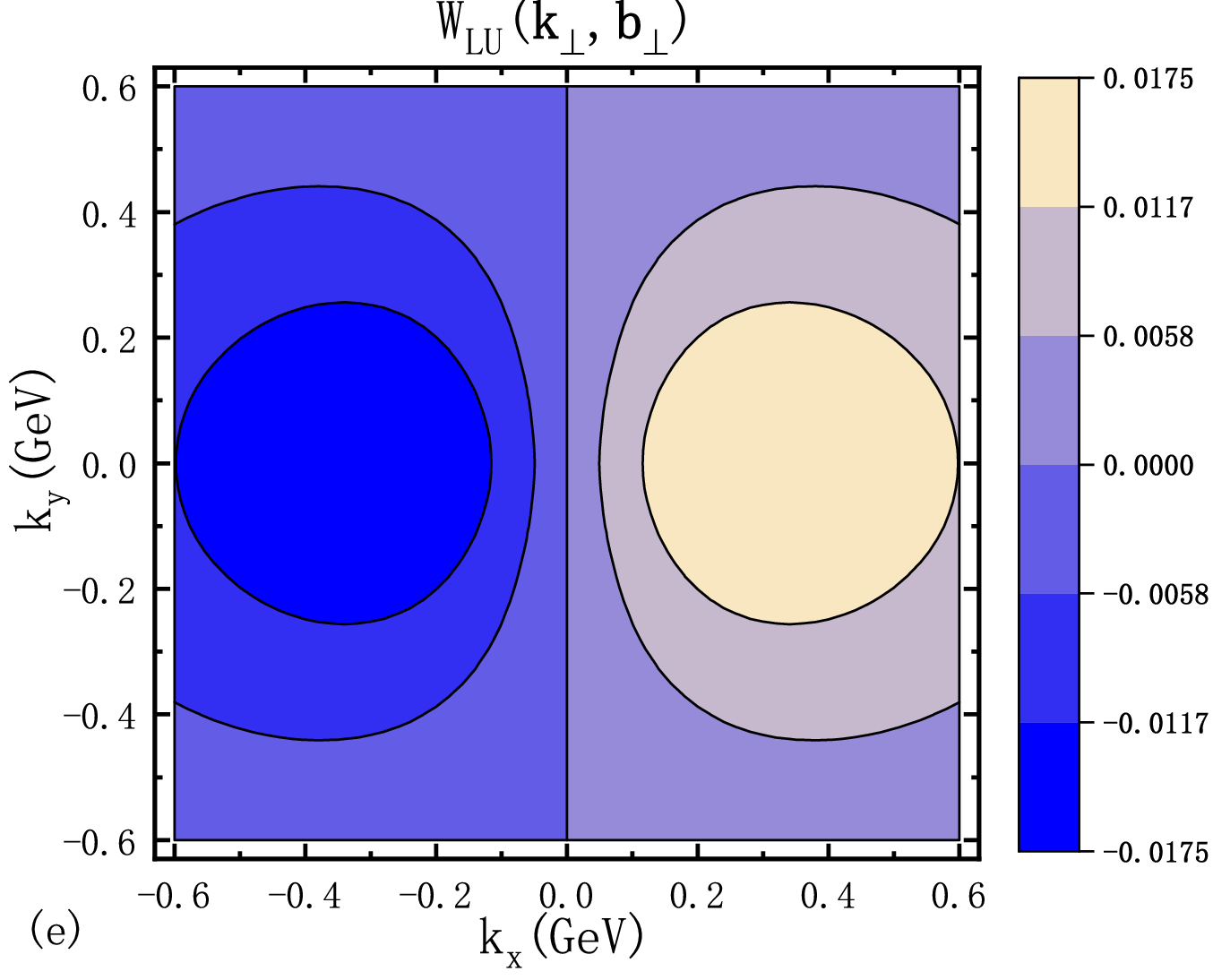}\\
  \includegraphics[width=0.42\columnwidth]{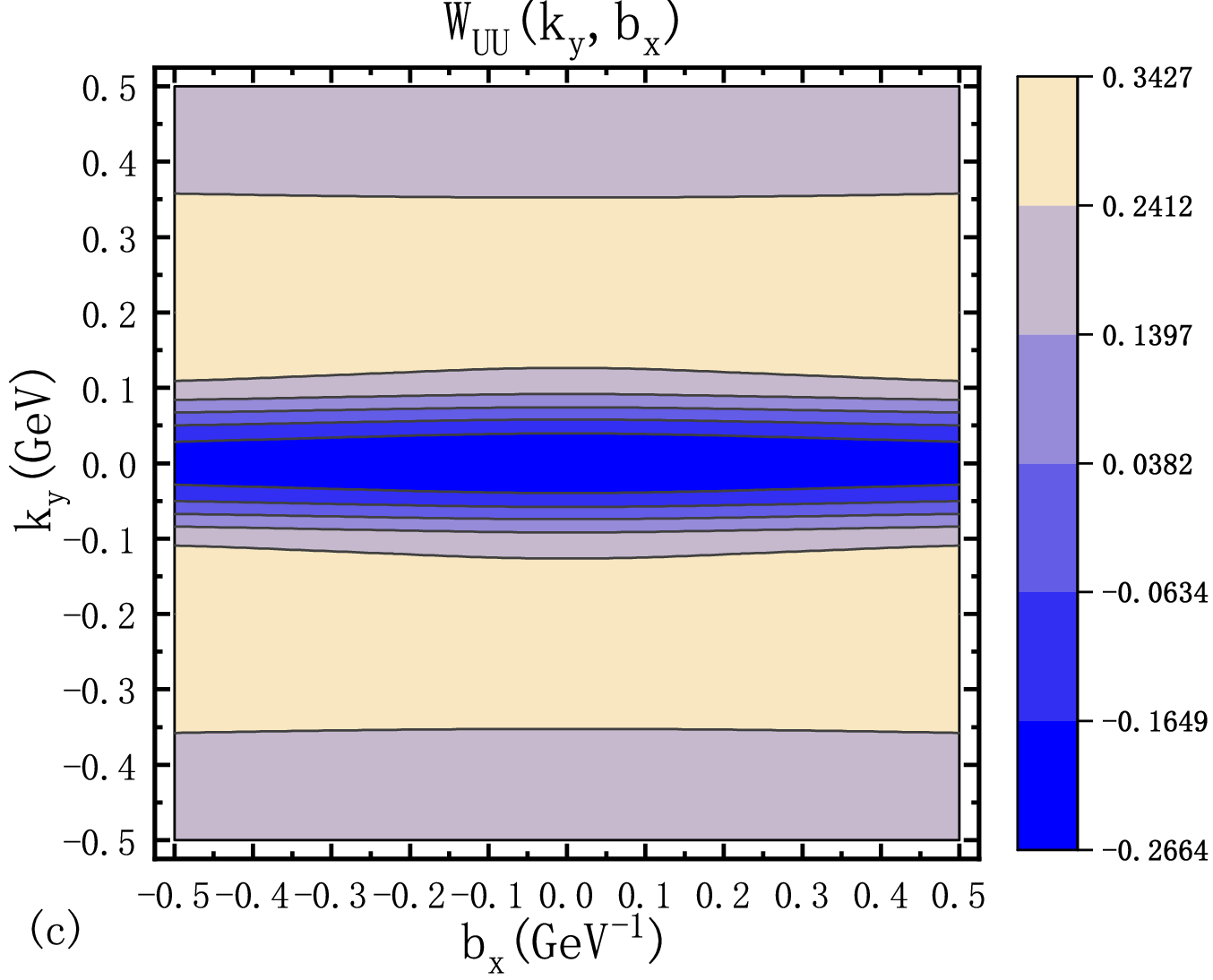}~~~
  \includegraphics[width=0.42\columnwidth]{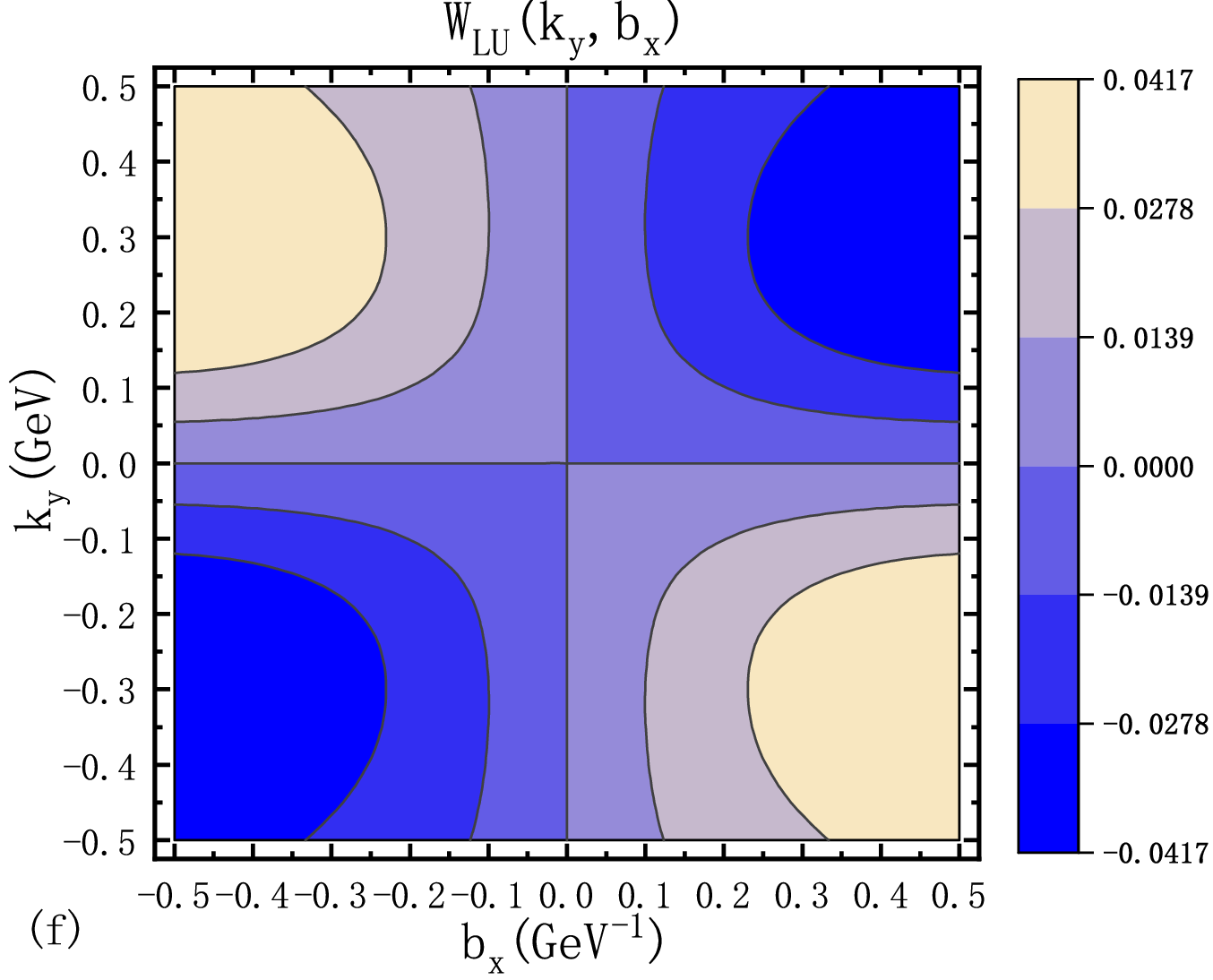}\\
\caption{The contour plots of the Wigner distributions $W_{UU}(\bm{k}_\perp,\bm{b}_\perp)$ (left panel) and $W_{LU}(\bm{k}_\perp,\bm{b}_\perp)$ (right panel). 
The first row displays the two distributions in $\bm{b}_\perp$ space with $\bm{k}_\perp=0.5 \ GeV \ \hat{\bm{e}}_y$. 
The second row displays the two distributions in $\bm{k}_\perp$ space with $\bm{b}_\perp=1 \ GeV^{-1} \ \hat{\bm{e}}_y$. 
The third row displays the two distributions in the mixed space of $b_x$ and $k_y$.}
\label{fig:wuu,wlu}
\end{figure}

In the left panel of Fig.~\ref{fig:wuu,wlu}, we plot the contour curves of $W_{UU}$, which 
is the Wigner distribution of the unpolarized gluon in an unpolarized proton. 
while in the right panel of Fig.~\ref{fig:wuu,wlu}, we plot the contour curves of $W_{LU}$.
The first row represents the Wigner distributions in $\bm b_\perp$ space with fixed $\bm{k}_\perp=0.5$ GeV  $\hat{\bm{e}}_y$. 
The second row shows the Wigner distributions in the $\bm k_\perp$ space with fixed $\bm{b}_\perp=1$ GeV$^{-1}\hat{\bm{e}}_y$.
The third row shows the Wigner distributions in the mixed $(b_x,k_y)$ space by integrating out $b_y$ and $k_x$. 
We find that the distribution in the $\bm b_\perp$ space is symmetric about $b_x$ and $b_y$ axis. 
However, it is not rotationally symmetric. 
Moreover, the peaks of the distribution are located in the regions $b_y\sim 1$ GeV$^{-1}$. 
This is different from the quark-target model calculation~\cite{More:2017zqp} in which the maximum magnitude of the distribution is in the $\bm b_\perp = \bm 0$ GeV$^{-1}$ region. 
In the $\bm{k}_\perp$ space, $W_{UU}$ is positive in the outer region and has a negative peak at the center. 
This is because $W_{UU}$ is a quasiprobabilistic distribution and may be negative in certain region. 
We verified that after integrating over $\bm b_\perp$, the results turn to be positive.
Besides, the distribution spreads more in the $x$ direction compared with the $y$ direction, which means that the probability of finding a gluon at large $k_x$ is higher than at large $k_y$. 
This asymmetry, which is due to the fact that $\bm{b}_\perp$ and $\bm{k}_\perp$ are chosen in the $y$ direction, indicates that the gluon here is more likely to have the configuration $\bm{k}_\perp \perp \bm{b}_\perp$ rather than $\bm{k}_\perp \ || \ \bm{b}_\perp$, just like that is observed for quarks in the light-cone constituent quark model~\cite{Lorce:2011kd}. 
The mixed plot shows the gluon probability density correlating $b_x$ and $k_y$, and this correlation is not restricted by the uncertainty principle. 
This density first increases from $k_y=0$ and then decreases. 
In the specific calculation, we find that $W_{UU}( k_y,b_x)$ has a negative minimum at $b_x= 0$ and $k_y=0$. 

As mentioned in Sec.~\ref{Sec:1}, the Wigner distribution can be used to calculate the expectation value of any single-particle physical observable $O$ from its phase-space average with the Wigner distribution as weighting factor:
\begin{align}
\langle O \rangle=\int dx d^2\bm{k}_\perp d^2\bm{b}_\perp O(x,\bm{k}_\perp,\bm{b}_\perp) W(x,\bm{k}_\perp,\bm{b}_\perp).
\end{align}
Then the average gluon OAM in the $z$ direction in a proton is expressed as
\begin{align}
\int dx d^2\bm{k}_\perp d^2\bm{b}_\perp (\bm{b}_\perp \times \bm{k}_\perp)_z W_{UU}(x,\bm{k}_\perp,\bm{b}_\perp)=0.
\label{eq:wuu OAM}
\end{align}
The result indicates that there is no net OAM for the unpolarized gluon in an unpolarized proton. 
The unpolarized nucleon does not possess a spin polarization, which means that the sum of all parton angular momentum is zero. 
Furthermore, due to the rotational invariance, the spin and the OAM of quarks and gluons can only be zero at the same time.

\begin{figure}
  \centering
  \includegraphics[width=0.42\columnwidth]{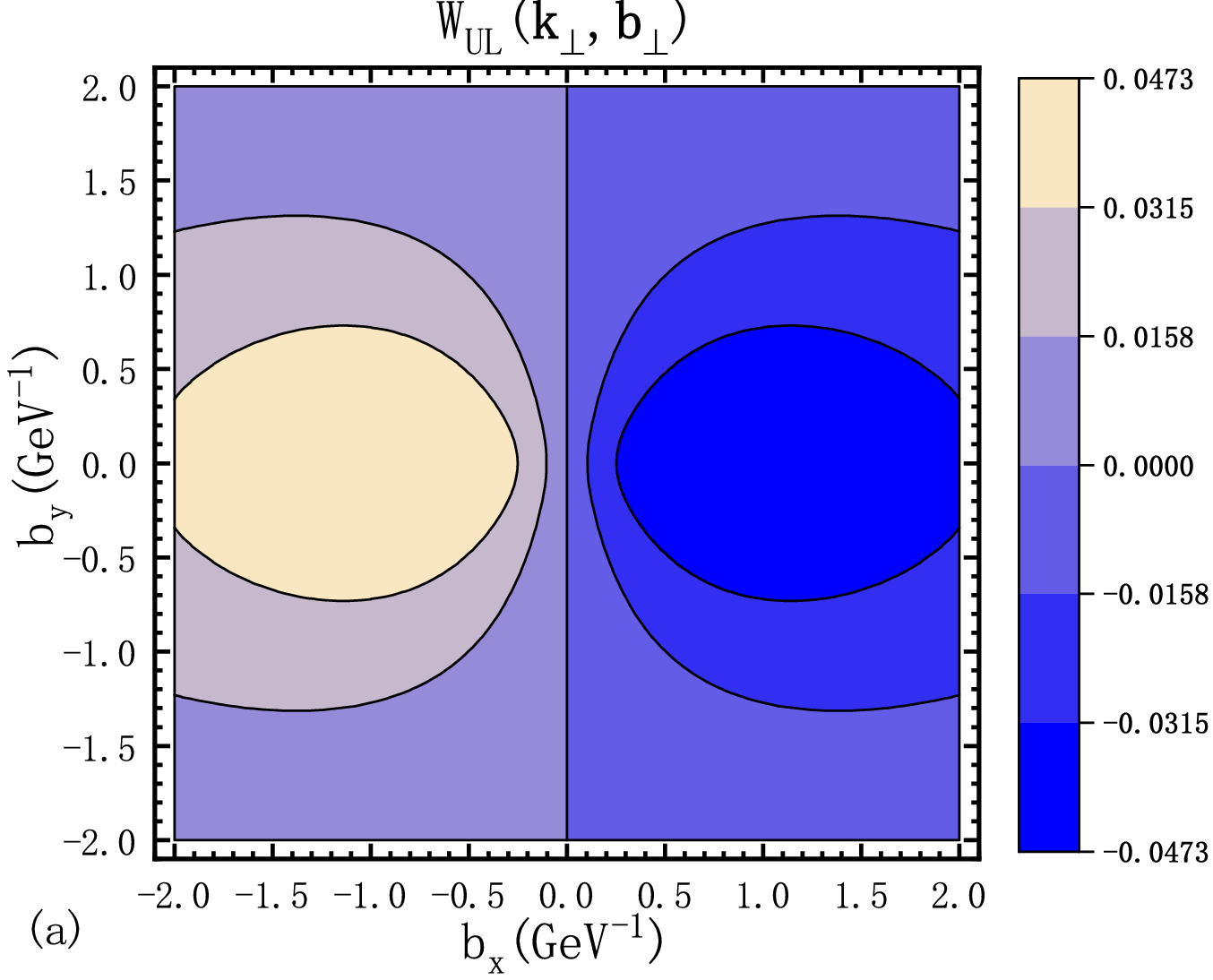}~~~
  \includegraphics[width=0.42\columnwidth]{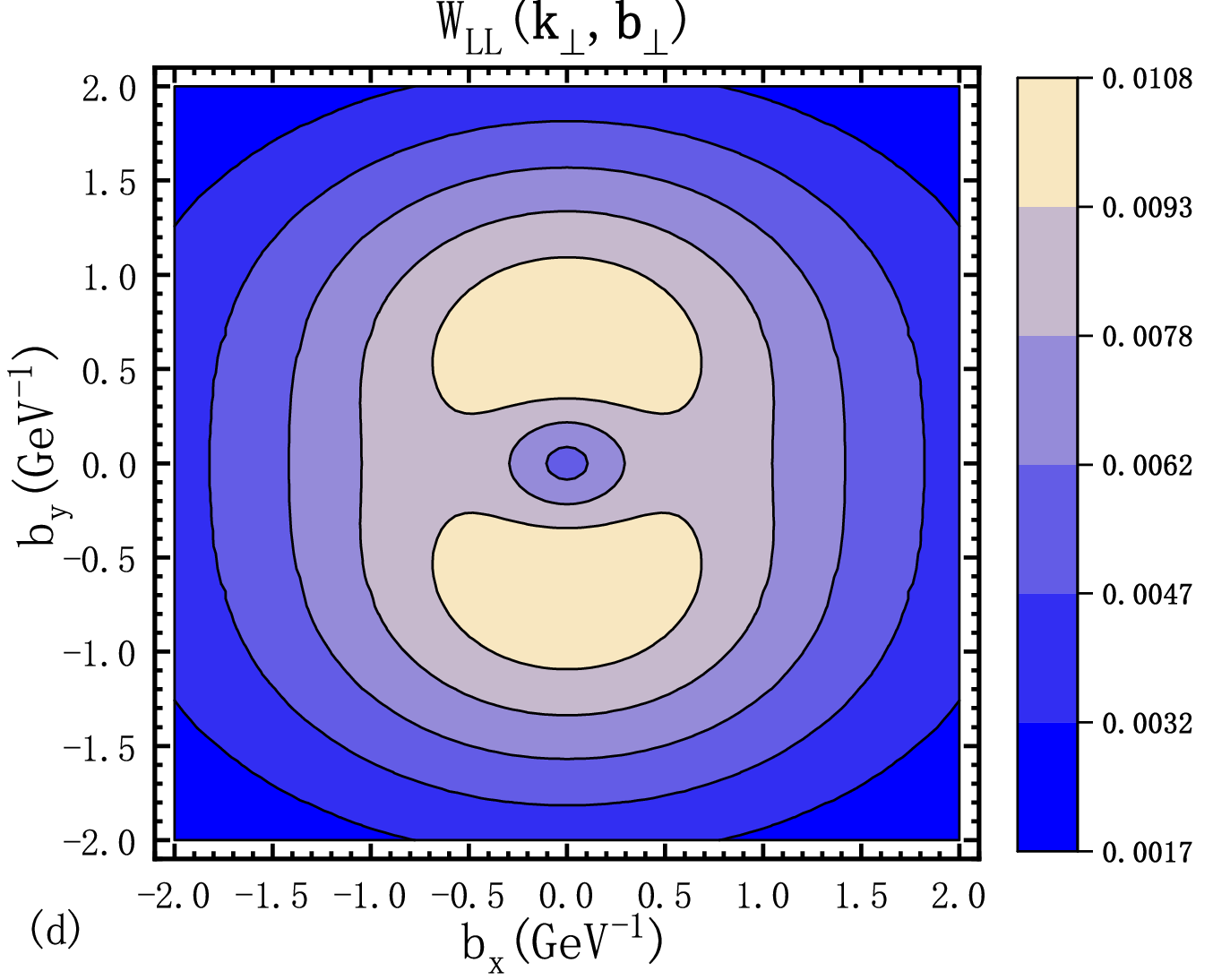}\\
  \includegraphics[width=0.42\columnwidth]{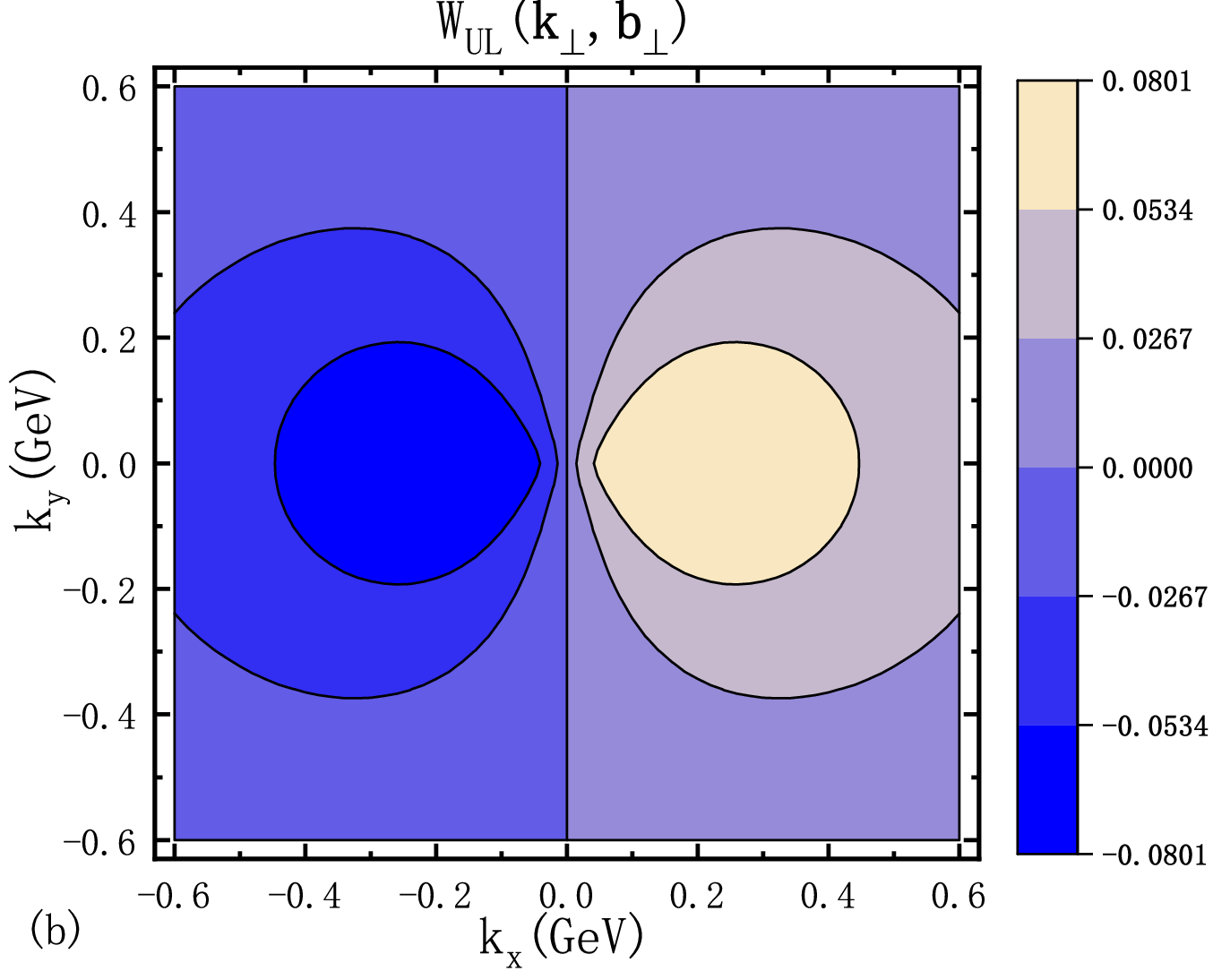}~~~
  \includegraphics[width=0.42\columnwidth]{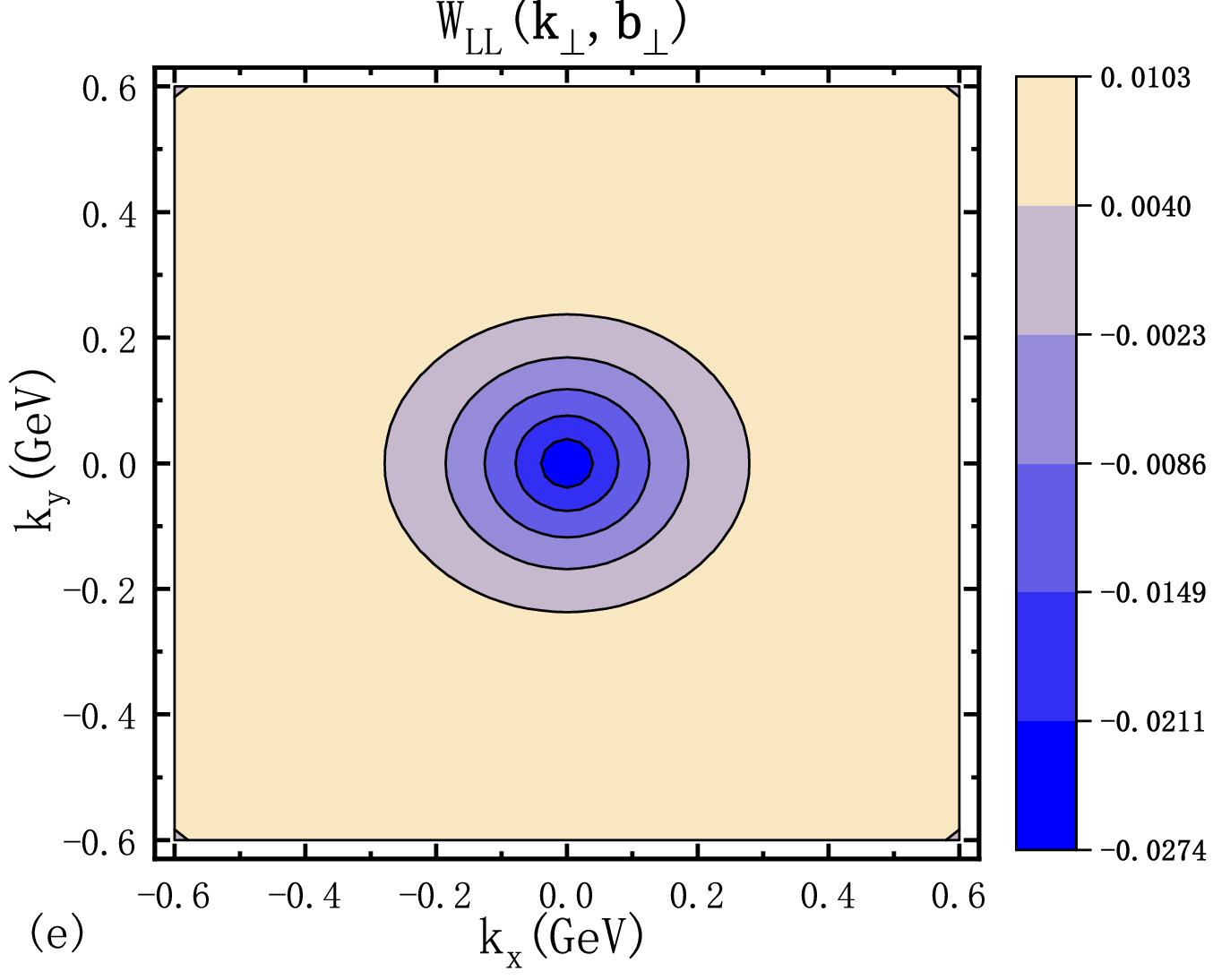}\\
  \includegraphics[width=0.42\columnwidth]{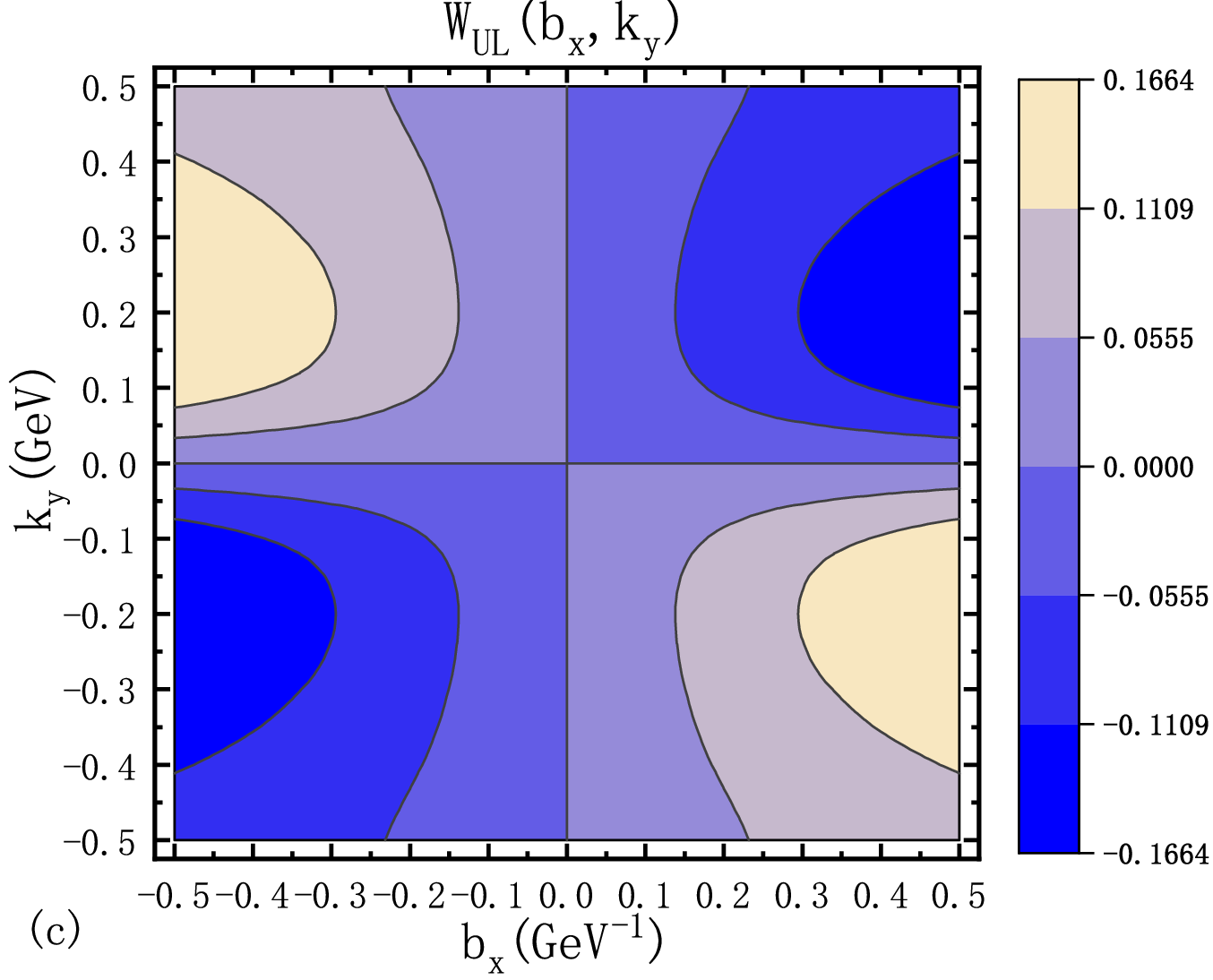}~~~
  \includegraphics[width=0.42\columnwidth]{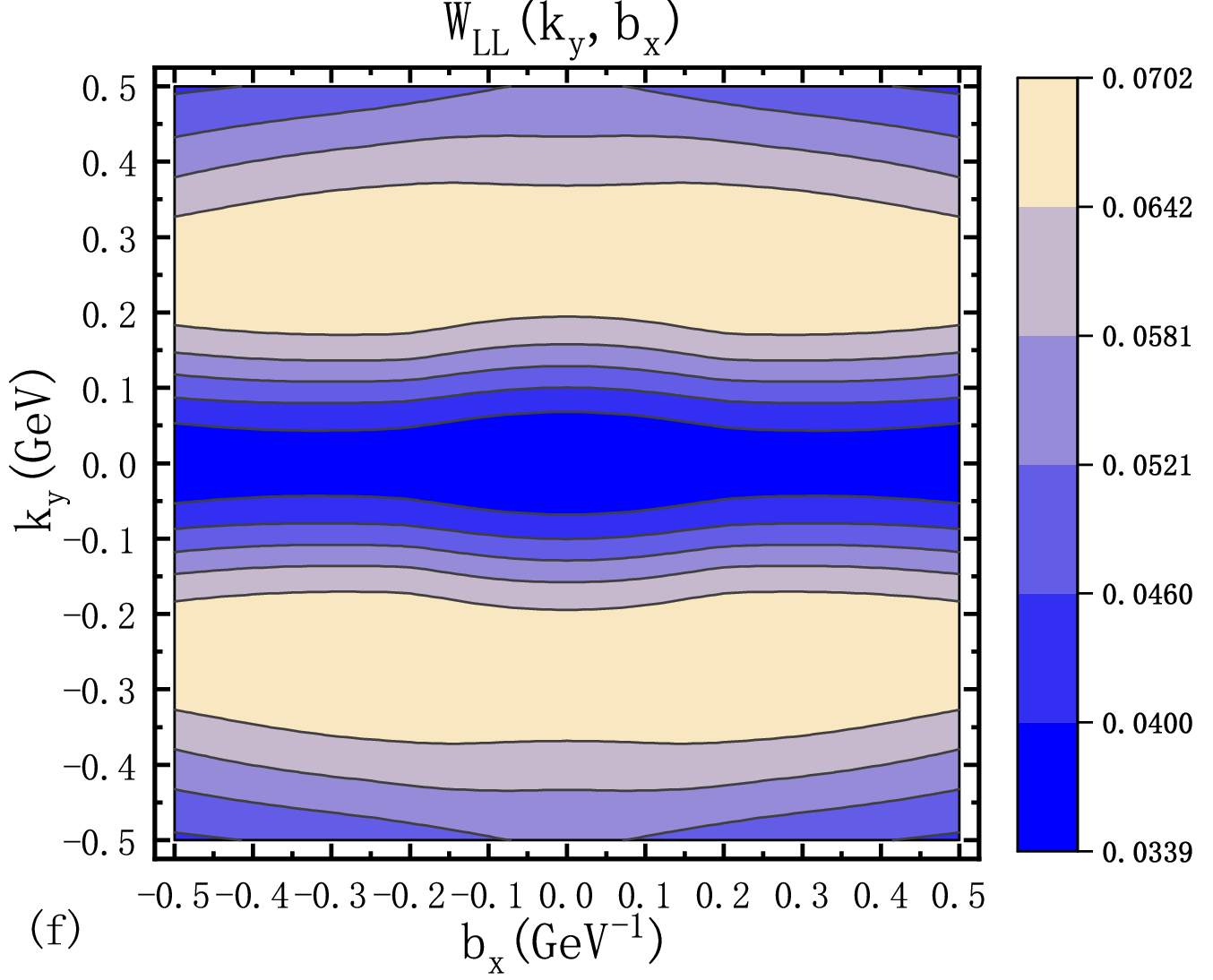}\\
  \caption{The contour plots of the Wigner distributions $W_{UL}(\bm{k}_\perp,\bm{b}_\perp)$ and $W_{LL}(\bm{k}_\perp,\bm{b}_\perp)$. The first row displays the two distributions in $\bm{b}_\perp$ space with $\bm{k}_\perp=k_\perp \ \hat{\bm{j}}=0.5 \ GeV \ \hat{\bm{j}}$. The second row displays the two distributions in $\bm{k}_\perp$ space with $\bm{b}_\perp=b_\perp \ \hat{\bm{j}}=0.5 \ GeV^{-1} \ \hat{\bm{j}}$. The third row displays the distributions in the mixed space of $b_x$ and $k_y$.}
  \label{fig:wul,wll}
\end{figure}

In the right panel of Fig.~\ref{fig:wuu,wlu}, we plot the contour curves of $W_{LU}$, which 
is the Wigner distribution of the unpolarized gluon in a longitudinally polarized proton. 
We observe a dipole structure both in $\bm{b}_\perp$ space and the $\bm{k}_\perp$ space.
The distribution is odd about the $b_y$ axis and $k_y$ axis. 
The positive (negative) peak in the $\bm{b}_\perp$ space appears in the $-b_x$ $(+b_x)$ region, while the case in $\bm{k}_\perp$ space is just the opposite. 
We also observe a quadrupole structure in the mixed space, where the negative peaks appear in the first and third quadrants and the positive peaks appear in the second and fourth quadrants. 
These multipole structures come from the fact that the factor $\epsilon_\perp^{ij}k_\perp^i\frac{\partial}{\partial b_\perp^j}$ in Eq.~(\ref{eq:wlu f14}) breaks the left-right symmetry of the plots. 
However, this asymmetry allows the existence of the non-zero net gluon OAM
\begin{align}
l_z^g=&\int dx d^2\bm{k}_\perp d^2\bm{b}_\perp (\bm{b}_\perp \times \bm{k}_\perp)_z W_{LU}(x,\bm{k}_\perp,\bm{b}_\perp)
\label{eq:wlu OAM}\\
=&-\int dx d^2\bm{k}_\perp \frac{\bm{k}_\perp^2}{M^2} F_{1,4}^g(x,0,\bm{k}^2_\perp,0,0),
\end{align}
which is known as the canonical OAM.

In the left panel of Fig~\ref{fig:wul,wll}, we show the Wigner distribution $W_{UL}$ which denotes the longitudinally polarized gluon in an unpolarized proton. 
The plots in $\bm{b}_\perp$ space and $\bm{k}_\perp$ space have a dipole structure, while the mixed plot in $b_x$-$k_y$ space has a quadrupole structure. 
Again, these multipole structures are due to the factor $\epsilon_\perp^{ij}k_\perp^i\frac{\partial}{\partial b_\perp^j}$ in Eq.~(\ref{eq:wul g11}) Here, the asymmetry from this factor leads to the gluon spin-orbit correlations
\begin{align}
C_z^g=&\int dx d^2\bm{k}_\perp d^2\bm{b}_\perp (\bm{b}_\perp \times \bm{k}_\perp)_z W_{UL}(x,\bm{k}_\perp,\bm{b}_\perp)\\
=&\int dx d^2\bm{k}_\perp \frac{\bm{k}_\perp^2}{M^2} G_{1,1}^g(x,0,\bm{k}^2_\perp,0,0).
\end{align}

In the right panel of Fig.~\ref{fig:wul,wll} we show the contour plots of $W_{LL}$, the longitudinally polarized gluon in a longitudinally polarized proton. 
In the $\bm{b}_\perp$ space, $W_{LL}$ is positive in the entire region. 
In the region $b_y=\pm 0.7$ GeV$^{-1}$ and $b_x=\pm 0$ GeV$^{-1}$, the distribution has the maximum magnitude.
In the $\bm{k}_\perp$ space, $W_{LL}$ is positive in the outer region and has a negative peak at the center. 
Both distributions and spread more in the $x$ direction than in the $y$ direction, indicating that the gluon also has the configuration $\bm{k}_\perp \perp \bm{b}_\perp$. 
The mixed plot shows that the probability of finding a gluon first increases with increasing $k_y$ and then decreases. 
We find that $W_{LL}(k_y,b_x)$ has a positive minimum at $b_x=0$ and $k_y=0$ in the model. 
In addition,
\begin{align}
\int dx d^2\bm{k}_\perp d^2\bm{b}_\perp (\bm{b}_\perp \times \bm{k}_\perp)_z W_{LL}(x,\bm{k}_\perp,\bm{b}_\perp)=0
\end{align}
indicates the isotropy of space like Eq.~(\ref{eq:wuu OAM}).

\begin{figure}
  \centering
  \includegraphics[width=0.45\columnwidth]{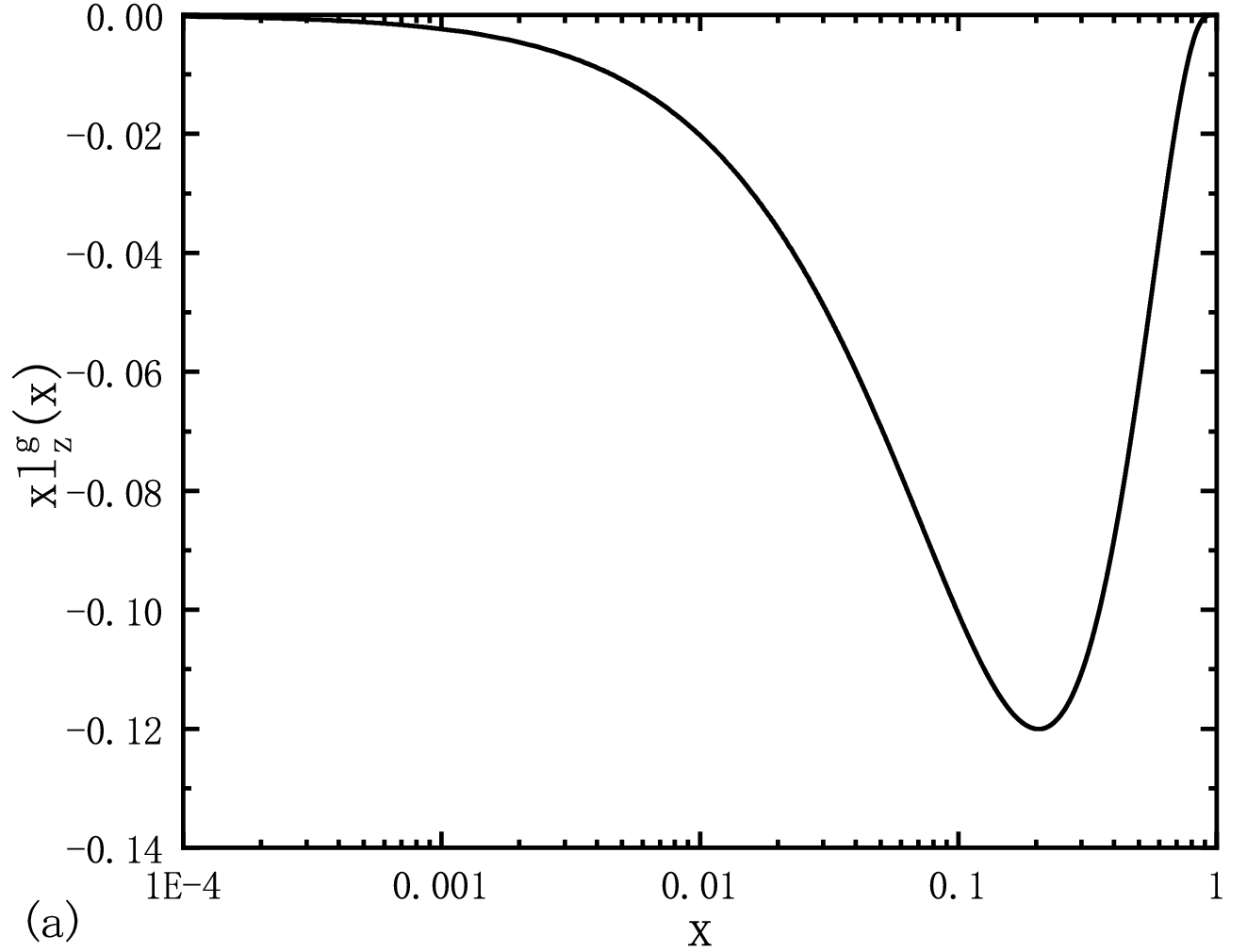}~~~
  \includegraphics[width=0.45\columnwidth]{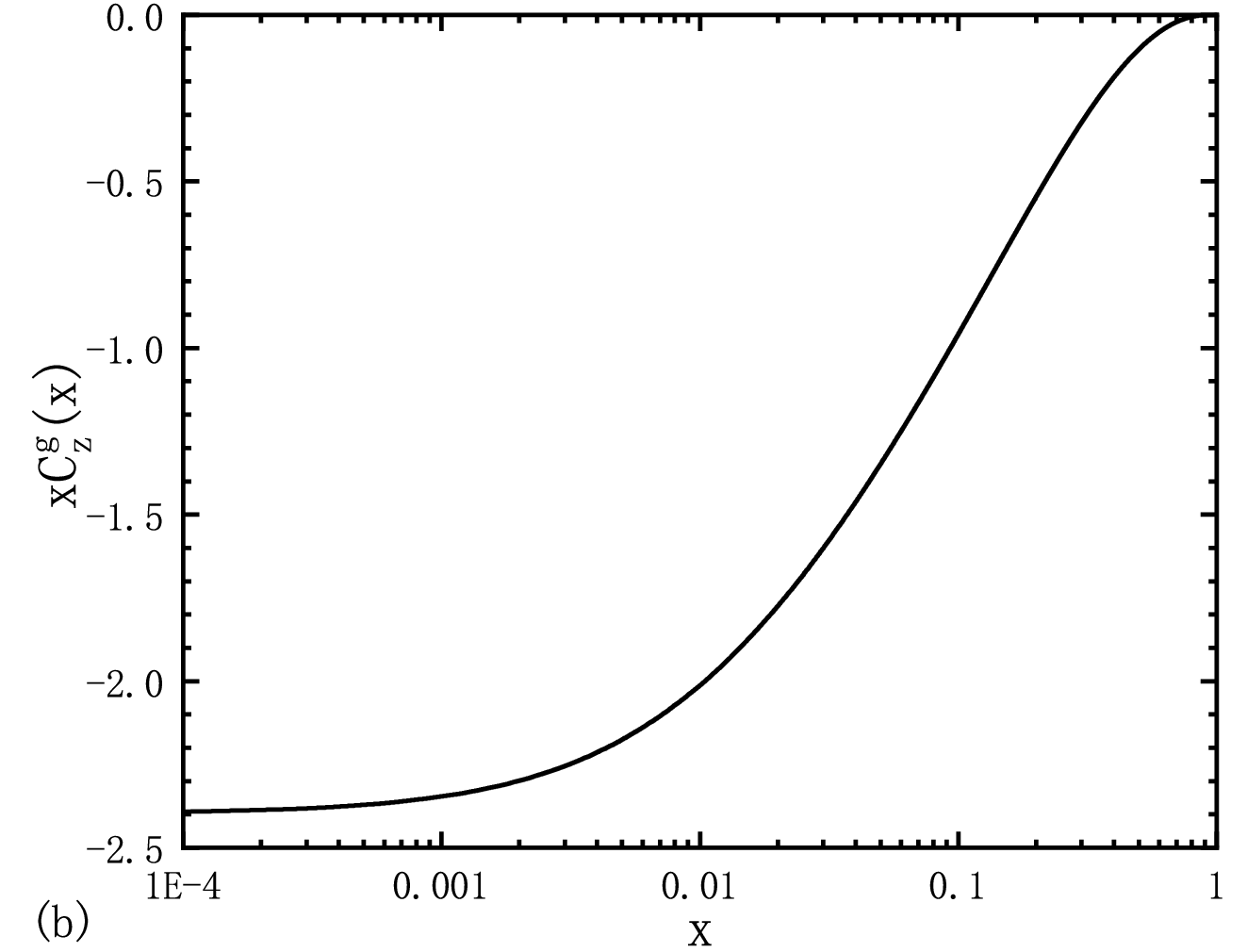}\\
  \caption{The left panel displays the dependence of the canonical gluon OAM $l_z^g(x)$ (timed with $x$) on $x$. The right panel displays the dependence of the gluon spin-orbit correlations $C_z^g(x)$ (timed with $x$) on $x$.}
  \label{fig:lzg,czg}
\end{figure}

Finally, we discuss the canonical gluon OAM and spin-orbit correlation of the gluon. Fig.~\ref{fig:lzg,czg}a shows the $x$-dependence of the canonical OAM $l_z^g(x)$ (timed with $x$). 
We observe that $xl_z^g(x)$ is negative and has a minimum value at the small $x$, which indicates that the contribution of $l_z^g(x)$ is mainly concentrated at the small-$x$ region. 
We also calculate the total gluon OAM and find $l_z^g \approx -0.333$. This result has a negative sign so that it reduces the total angular momentum contribution of the gluon to the proton spin. 
Fig.~\ref{fig:lzg,czg}b shows the dependence of the spin-orbit correlations $C_z^g(x)$ (timed with $x$) on $x$. 
Unlike the case of quarks~\cite{Mukherjee:2014nya,Mukkherjee:2015phf}, due to $F_{1,4}^g\neq G_{1,1}^g$, $l_z^g$ and $C_z^g$ have different result in this model. However, we notice that $\l_z^g(x) \approx xC_z^g(x)$ with $x \rightarrow 0$. 
We also find that $xC_z^g(x)$ is negative. This means that the gluon spin and OAM tend to be antialigned, which is as same as the case of quarks~\cite{Lorce:2014mxa,Tan:2021osk}. Finally, $xC_z^g(x)$ has a minimum value at the very small $x$, which indicates that the main contribution of $C_z^g(x)$ is also concentrated at the small-$x$ region.

\section{Conclusion}\label{Sec:5}

In this work, we investigated the Wigner distributions of gluon in an unpolarized proton and in a longitudinally polarized proton.  
In the study we applied a light-cone spectator model, in which the proton target is regarded as a two-body composite system composed of an active gluon and a spectator particle (uud). 
The four Wigner distributions $W_{UU}$, $W_{LU}$, $W_{UL}$ and $W_{LL}$ as well as the GTMDs were defined from the gluon generalized correlator at $\xi=0$. 
Using the light-cone wave functions, we obtained the expressions of the Wigner distributions and GTMDs within the overlap representation. 
We performed numerical calculations and presented the contour plots of the gluon Wigner distributions $W_{UU}$, $W_{LU}$, $W_{UL}$ and $W_{LL}$ in the transverse position space, in the transverse momentum space and in the mixed space, respectively. 
We find that $W_{UU}$ and $W_{LL}$ are positive in the $b_\perp$ space, while they are negative in the small $k_\perp$ region are similar, and the gluons described by these two distributions have the configuration $\bm{k}_\perp \perp \bm{b}_\perp$. 
The corresponding plots of $W_{LU}$ and $W_{UL}$ shows the presence of the multi-pole structures, which imply the existence of the canonical gluon OAM and spin-orbit correlations, respectively. 
In our model, the canonical OAM is different from the spin-orbit correlations. 
We found that the main contributions of $l_z^g(x)$ and $C_z^g(x)$ are both concentrated in the small-$x$ region. 
In addition, $l_z^g(x)$ and $C_z^g(x)$ are both negative, and the former implies that the total gluon OAM will reduce the total angular momentum contribution of the gluon to the proton, while the later indicates that the gluon spin and OAM are antialigned.

\section*{Acknowledgements}
This work is partially supported by the National Natural Science Foundation of China under grant number 12150013.

\end{document}